\documentclass[a4paper,11pt]{article}
\pdfoutput=1
\usepackage[usenames,dvipsnames]{color}

\usepackage{jcappub}

\usepackage{listings}
\usepackage{color}

\usepackage{algorithm}
\usepackage[noend]{algpseudocode}

\makeatletter
\def\BState{\State\hskip-\ALG@thistlm}
\makeatother

\definecolor{dkgreen}{rgb}{0,0.6,0}
\definecolor{gray}{rgb}{0.5,0.5,0.5}
\definecolor{mauve}{rgb}{0.58,0,0.82}

\newcommand{\code}[1]{\texttt{#1}\xspace}

\providecommand{\e}[1]{\ensuremath{\times 10^{#1}}}
\def\mpl{M_\mathrm{Pl}}
\def\impc{\; \mathrm{Mpc}^{-1}}
\def\fnl{f_\mathrm{NL}}
\def\taunl{\tau_\mathrm{NL}}

\newcommand{\order}[1]{\ensuremath{\mathcal{O}(#1)}}

\usepackage{xspace}
\newcommand{\MMC}{\textsc{MultiModeCode}\xspace}

\newcommand{\ud}{\mathrm{d}}
\newcommand{\pd}{\partial}

\def\bk{{\mathbf{k}}}
\def\f {{\phi}}
\def\df {\dot \phi_0}

\def\calR{{\cal R}}

\def\calS{{\cal S}}

\def\PIJ{{\mathcal{P}_{\delta \phi}^{IJ}}}
\def\Pten{{\mathcal P_h}}
\def\Pent{{\mathcal P_\mathrm{ent}}}
\def\Ppnad{{\mathcal P_{\delta P, \mathrm{nad}}}}
\def\Cross{{\mathcal P_\mathrm{\calR \calS}}}
\def\PR{{\mathcal P_\calR}}
\def\PS{{\mathcal P_\calS}}
\def\Pp{{\mathcal P_{\delta P}}}

\def\Ph{{\mathcal P_{h}}}

\def\dPnad{\delta P_\mathrm{nad}}

\renewcommand{\th}{{\ensuremath{^\mathrm{th}}}}

\title{MultiModeCode: An efficient numerical solver for multifield inflation}

\author[1]{Layne C. Price,}
\author[2,3]{Jonathan Frazer,}
\author[4]{Jiajun Xu,}
\author[5]{Hiranya V. Peiris,}
\author[1]{and Richard Easther}

\affiliation[1]{Department of Physics,  University of Auckland \\ Private Bag 92019,  Auckland, New Zealand}
\affiliation[2]{Department of Theoretical Physics, University of the Basque Country, UPV/EHU \\ 48040 Bilbao, Spain}
\affiliation[3]{IKERBASQUE, Basque Foundation for Science \\ 48011 Bilbao, Spain}
\affiliation[4]{Department of Physics, University of Wisconsin-Madison \\ Madison, WI 53706, USA}
\affiliation[5]{Department of Physics and Astronomy,  University College London \\ London WC1E 6BT, UK}

\emailAdd{lpri691@aucklanduni.ac.nz}
\emailAdd{j.frazer@ucl.ac.uk}
\emailAdd{jiajun.xu@gmail.com}
\emailAdd{h.peiris@ucl.ac.uk}
\emailAdd{r.easther@auckland.ac.nz}

\begin{document}

\abstract{
  We present \MMC,\footnote{Available at \url{www.modecode.org}.} a Fortran 95/2000 package for the numerical exploration of multifield inflation models. This program facilitates efficient Monte Carlo sampling of prior probabilities for inflationary model parameters and initial conditions and is the first publicly available code that can efficiently generate large sample-sets for inflation models with $\mathcal O(100)$ fields. The code numerically solves the equations of motion for the background and first-order perturbations of multi-field inflation models with canonical kinetic terms and arbitrary potentials, providing the adiabatic, isocurvature, and tensor power spectra at the end of inflation. For models with sum-separable potentials \MMC also computes the slow-roll prediction via the $\delta N$ formalism for easy model exploration and validation.  We pay particular attention to the isocurvature perturbations as the system approaches the adiabatic limit, showing how to avoid numerical instabilities that affect some other approaches to this problem. We demonstrate the use of \MMC by exploring a few toy models. Finally, we  give a concise review of multifield perturbation theory and a user's manual for the program.
}

\maketitle
\flushbottom

\section{Introduction}

Many simple models of inflation  adeptly  reproduce the observed properties of the primordial cosmological perturbations  \cite{Hinshaw:2012aka,Ade:2013xsa,Ade:2013zuv,Ade:2013uln}, predicting a nearly scale-invariant power spectrum and minimal amounts of primordial non-Gaussianity.  In the slow-roll, single-field paradigm the predictions of a given model are easily determined as an algebraic function of the field's potential $V$ and its derivatives in terms of a hierarchy of slow-roll parameters.  The resulting observables are simple to compute and easy to interpret.

However, relaxing any of the basic assumptions of the slow-roll, single-field models complicates this simple analysis.  In particular, for many inflationary scenarios (\emph{e.g.}, multifield inflation, gauge inflation, and non-minimal couplings),
the background and mode equations are complex systems of coupled, nonlinear ODEs, making analysis difficult in all but a few cases.
Furthermore, while slow-roll, single-field inflation is a simple and easily understood model, it may not necessarily be considered natural in the context of high-energy theories.  For example, low energy effective theories derived from string theory generically contain hundreds of scalar fields with complicated interactions, and many theories consider non-minimal couplings to the Ricci scalar (for a recent review, see Ref.~\cite{Baumann:2014nda}). While analytical studies have been able to overcome subsets of these problems, most of the techniques that have been used are situation-specific, which limits their applicability to novel models.

While significant progress can be made in the slow-roll limit, only  numerical techniques can explore the full predictions of more complex inflation models.  Even in the purely homogeneous limit, numerically solving the nonlinear Klein--Gordon equation for the homogeneous background fields reveals many interesting features that do not arise in slow-roll analyses, \emph{e.g.}, sensitivity to initial conditions \cite{Easther:1997hm,Clesse:2009ur,Easther:2013bga,Easther:2014zga}.  These complications lead naturally to the numerical exploration of inflationary models.

In this paper we present and describe \MMC,\footnote{Publicly available at \url{www.modecode.org}.} an efficient Fortran 95/2000 package that numerically solves the equations of motion for the background fields and the first-order perturbations for multifield inflation models in which the fields have canonical kinetic terms and are minimally coupled to gravity.  \MMC\ calculates the adiabatic, tensor, and various isocurvature power spectra as a function of scale $k$, but does not evaluate higher order correlators.  If the potential is sum-separable, \MMC\ uses the solution to the background equations of motion to evaluate the slow-roll $\delta N$ predictions for the scalar and tensor power spectra and their derivatives near the pivot scale $k_*$, also giving the slow-roll results for $n_s$, $r$, $f_\mathrm{NL}$, etc.  The code has been extensively tested with various compilers, including the open-source GNU Fortran compiler.

Several numerical codes have been developed to study single-field models~\cite{Adams:2001vc,Peiris:2003ff,Martin:2006rs, Hall:2007qw,Bean:2007eh,Lorenz:2007ze,Martin:2010kz,Martin:2010hh,Martin:2013nzq}.  Here, we build on \textsc{ModeCode}~\cite{Mortonson:2010er,Easther:2011yq,Norena:2012rs}, which was developed to test single-field inflation models and interfaced with tools such as \textsc{CAMB}~\cite{Lewis:1999bs}, \textsc{CosmoMC}~\cite{Lewis:2002ah}, and \textsc{MultiNest}~\cite{Feroz:2008xx}.  \textsc{ModeCode} was designed for the Bayesian analysis of inflation and  used by the \emph{Planck} collaboration~\cite{Ade:2013uln} to obtain the posterior probabilities and marginal likelihoods for inflation models.  Moving to the multifield case significantly increases the numerical demands on the solver, and puts a premium on efficiency due to the much greater computational resources required by these analyses.  A few codes exist to analyze multified models, but the publicly available codes are inadequate for models with many fields and arbitrary potentials.  Notably, \textsc{Pyflation} \cite{Huston:2009ac,Huston:2011fr,Huston:2011vt,Huston:2013kgl} is an object-oriented Python code that uses the same method we employ here for solving the perturbation equations, but cannot easily generate large samples due to the speed constraints imposed by a dynamic programming language.

This significant extension to \textsc{ModeCode} can be used to study the power spectra of analytically intractable multifield inflationary potentials, and to explore the generic predictions of complex models by marginalizing over large numbers of possible parameters.  Complementing currently available codes~\cite{Huston:2009ac,Huston:2011fr,Huston:2011vt,Huston:2013kgl}, \MMC specializes in obtaining large Monte Carlo samples of initial conditions and parameter prior probabilities.  To help users familiarise themselves with \MMC\ the package includes initial conditions priors  used in Refs.~\cite{Frazer:2013zoa,Easther:2013bga,Easther:2013rva}.  The ability of this code to efficiently generate large Monte Carlo samples has permitted studies of the generic predictions of multifield inflation models with more than 100 fields~\cite{Easther:2013rva,Price:2014ufa}.

In practice, the code can simulate the evolution of the mode equations for $\mathcal O(10^2)$ fields,\footnote{Estimates regarding field number are based on $N_{f}$-quadratic inflation, which is not numerically intensive.} but will become inefficient for significantly more fields due to the increasing dimensionality of the system, which increases with the number of fields as $\mathcal O(N_f^2)$.  However, it can efficiently sample the evolution of the background equations of motion for at least $\mathcal O(10^3)$ fields.  While solving just the background equations allows the exploration of background dynamics for such a large number of fields, if the model is sum-separable, then it will also give the slow-roll predictions for the adiabatic curvature power spectrum, as well as $\fnl$ and $\taunl$, in terms of the $\delta N$ approximation. This should be valid when the fields are much lighter than $H$ at horizon crossing and slow-roll holds throughout the duration of inflation.  \MMC is released with several example models already implemented and it is straightforward to add to this number.

In \S\ref{sect:results}, we demonstrate the features of \MMC with an $N_f$--flation potential with a sharp step, which we parametrize by a hyperbolic tangent function, following Refs.~\cite{Adams:1997de,Adams:2001vc}.  We show that, in addition to oscillatory features in the adiabatic curvature power spectrum that are expected from the single-field analysis~\cite{Adams:2001vc,Adshead:2011jq}, with more than one field there are also oscillatory features in the isocurvature spectra, which might result in non-trivial evolution of the power spectrum after inflation.
We also show that the numerical computation of isocurvature modes results in an inherent numerical instability, since some definitions of isocurvature perturbations involve computing the difference between two quantities that are of the same order of magnitude.  This induces a dominant numerical error when these two quantities begin to approach the adiabatic limit.  We overcome this problem by implementing a modified definition of isocurvature perturbations~\cite{Easther:2013rva}, which is numerically stable to many more orders of magnitude than some alternative definitions. We also implement a geometrical optics indicator of isocurvature evolution as first presented in Ref.~\cite{Seery:2012vj}. While this measure only relies on background quantities and also does not suffer from instabilities, as implemented here it does not provide an absolute value of isocurvature, only an indicator of its growth or decay.

Finally, in \S\ref{sect:pert_review} we provide a concise review of multifield perturbation theory with the aim of dispelling misconceptions that exist about this topic, which the enlightened reader can skip.

\section{Features of \MMC}

We begin by highlighting some of the useful characteristics of \MMC.

\paragraph{Speed:}
The purpose of \MMC is to provide a fast and efficient solver that is well-tested and can be applied to a wide range of possible inflationary scenarios.  \MMC is written in Fortran 95/2000, increasing its capabilities relative to existing codes~\cite{Huston:2009ac,Huston:2011fr,Huston:2011vt,Huston:2013kgl} and making it tractable to investigate models with many fields or to obtain large Monte Carlo samples from a model's parameter space.  In particular, prototype versions of this program were used in Refs.~\cite{Easther:2013rva,Price:2014ufa} to analyze large samples of 100-field $N_f$-monomial inflation.

\paragraph{Generality:}
The code facilitates Bayesian approaches to studying inflation, where the model's parameters are drawn from prior probabilities from which we can compute a probability distribution for specified observable associated with the model.  We consider simple situations, \emph{e.g.}, evolving a  model given fixed model parameters and initial conditions, as sub-cases of the more general Bayesian framework.  To facilitate the use of general priors we have implemented the sampling routines in modules which are simple to adapt and restructure for the user's purposes.

\paragraph{Robustness:}
The program exits gracefully when encountering fatal errors of either  a technical or cosmological nature, while also catching specific errors that might only affect one particular configuration of the model.  We have extensively checked the program output on various Macintosh and Linux machines with both the \textsc{gfortran} and \textsc{ifort} compilers, and include both a fourth-order Runge-Kutta integrator and an implicit backward-difference formula method, which is suitable for stiff problems.

\paragraph{Statistics:}
\MMC provides  pivot-scale observables,  summarized in Table~\ref{table:observables} and can sample the adiabatic and isocurvature power spectra as a function of scale $k$.  We have implemented a variety of numerically stable indicators of the amount of isocurvature present in the system.

\paragraph{Slow-roll comparison:}
If the potential $V$ is sum-separable,  \MMC can also calculate observables using the $\delta N$ approximation, which assumes slow-roll.  Since these quantities rely only on  solutions of the background equations of motion they are efficient and simple to calculate, scaling with the number of fields as $\mathcal O(N_f)$.  Consequently, if the model is well-described by the slow-roll approximation between horizon crossing and the end of inflation, computing observables in the $\delta N$ formalism is efficient and easy.

\begin{table}
  \centering
  \begin{tabular}{  l  p{7cm} l }
    \hline
    \hline
    Power spectra (PS)  &  Type  & Reference \\
    \hline
    $\PR(k)$   \dotfill  &
      Adiabatic scalar spectrum \dotfill &
    Eq.~\eqref{eqn:pad} \\
    $\PS(k)$   \dotfill  &
      Isocurvature spectrum \dotfill &
         Eq.~\eqref{eqn:piso}  \\
    $\Ppnad(k)$\dotfill  &
      Non-adiabatic pressure spectrum \dotfill &
      Eq.~\eqref{eqn:ppnad} \\
    $\Pent(k)$ \dotfill  &
      Entropic spectrum \dotfill &
      Eq.~\eqref{eqn:pent} \\
    $\Cross(k)$\dotfill  &
      Adiabatic--non-adiab. cross spectrum \dotfill &
      Eq.~\eqref{eqn:cross} \\
    $\mathcal P_h(k)$ \dotfill  &
      Tensor spectrum \dotfill &
      -- \\
    \hline
    \hline
    Observable at $k_*$ &  Name  & Description \\
    \hline
    $A_s$ \dotfill &
         Scalar amplitude \dotfill&
         $\PR(k_*)$  \\
    $A_\mathrm{iso}$   \dotfill &
         Isocurvature ampl. \dotfill &
                     $\PS(k_*)$ \\
    $A_\mathrm{Pnad}$  \dotfill &
       Non-adiab. pressure ampl. \dotfill &
                  $\Ppnad(k_*)$ \\
    $A_\mathrm{ent}$   \dotfill &
               Entropy ampl. \dotfill &
                   $\Pent(k_*)$ \\
    $A_\mathrm{Cross}$ \dotfill &
             Cross spectra ampl. \dotfill &
                  $\Cross(k_*)$  \\
    \hline

    $n_s$  \dotfill            &
                 Scalar spectral index  \dotfill &
      $\mathcal D_* \log \PR+1$ \\
    $n_t$  \dotfill            &
          Tensor spectral index  \dotfill &
      $\mathcal D_* \log \Pten$ \\
    $n_\mathrm{iso}$  \dotfill &
                 Isocurvature spectral index  \dotfill &
        $\mathcal D_* \log \PS$ \\
    $n_\mathrm{ent}$  \dotfill &
                  Entropy spectral index  \dotfill &
      $\mathcal D_* \log \Pent$ \\
    $n_\mathrm{Pnad}$ \dotfill &
      Non-adiab. pressure spectral index  \dotfill &
     $\mathcal D_* \log \Ppnad$ \\
    \hline

    $\alpha_s$  \dotfill     &
            Scalar running \dotfill &
   $\mathcal D_*^2 \log \PR$ \\
    $r$    \dotfill          &
      Tensor-to-scalar ampl. \dotfill &
       $\Pten(k_*)/\PR(k_*)$ \\
    $\Theta$ \dotfill        &
         Bundle width  \dotfill &
       Eq.~\eqref{eqn:Theta}  \\
    $\cos \Delta$ \dotfill   &
  $\omega$-$s$ correlation angle \dotfill &
                 Eq.~\eqref{eqn:Delta} \\
    \hline
    \hline
  \end{tabular}

  \caption{Typical observables at the pivot scale $k_*$.  The derivative $\mathcal D_* \equiv \ud/\ud \log k$ is evaluated at $k=k_*$.  \MMC can also generate the full power spectra as a function of scale $\mathcal P(k)$.}
  \label{table:observables}
\end{table}

\section{A brief review of multifield perturbation theory}
\label{sect:pert_review}

We begin with a short review of first-order, non-interacting multifield perturbation theory before describing \MMC and the dynamics of many-field inflation.  There are some substantial differences between single-field and multifield inflation, which we highlight in Section~\ref{ssect:highlight}.  Table~\ref{table:observables} gives a list of the pivot-scale observables that \MMC computes.

There are a few excellent reviews of this topic~\cite{Lyth:1998xn,Langlois:2008mn,Lyth:2009zz,Huston:2011fr} and we particularly recommend Refs.~\cite{Salopek:1988qh,Bassett:2005xm} for more information.  We first present the nuts-and-bolts of the mode function approach to first-order, multifield perturbations, which is implemented in \MMC. Then we describe the widely-used $\delta N$-formalism, which has also been implemented for ease of use and for comparison to the perturbation solutions.

\subsection{The highlights}
\label{ssect:highlight}

Multifield inflation  differs from the single field case in the following important respects.

\paragraph{Isocurvature:}
Multifield inflation generally permits both adiabatic and isocurvature perturbations.
Adiabatic perturbations are related by a gauge transformation to the curvature perturbation on comoving hypersurfaces $\mathcal R$, while isocurvature perturbations are \emph{entropic} perturbations between different matter components on flat hypersurfaces.  In single-field inflation there is only one matter component, so there are only adiabatic perturbations.

\paragraph{Super-horizon evolution:}
Isocurvature perturbations source adiabatic perturbations, causing them to evolve even on super-horizon scales. While this can generate novel signatures such as non-Gaussianity, this can also be problematic for comparing the predictions of a model with observation: unless isocurvature modes decay into an \emph{adiabatic limit} before the end of inflation, the curvature perturbation does not become conserved and is thus sensitive to post-inflationary physics.

\paragraph{The two-index mode function:}
With more than one field, either (a) the direct interaction between fields or (b) the gravity-mediated interaction will mix the particle creation and annihilation operators as a function of time \cite{Salopek:1988qh}.  Instead of a single index mode function, we therefore need to solve for a mode matrix $\psi_{IJ}$, where $\delta \phi_I = {\psi_{IJ}} a^J$, for $N_f$ annihilation operators $a^J$.

\paragraph{Initial conditions dependence:}
Multifield inflation models have an infinite number of possible inflationary solutions each of which can, in principle yield a different perturbation spectrum. Consequently, the observable spectra for multifield models can depend on their initial conditions in ways that have no direct analogue in slow-roll, single-field models, which have only one possible trajectory in field-space.

\paragraph{Inherently stochastic predictions:}
Even if the potential $V$ is completely fixed, multifield models will give an inherent spread of predictions due to the allowed variance in the fields' initial conditions.  In general, multifield models will predict a variety of spectra, unless the stochasticity in the initial conditions can be controlled \emph{a priori}.

\subsection{Classical background}

Consider $N_f$ scalar fields $\phi_I$ with the matter sector of the action given by
\begin{equation}
  S = \int d^4x \sqrt{-g} \left[ - \frac{1}{2} \partial_\mu \phi_I \partial^\mu \phi^I - V(\phi_I)  \right]  ,
  \label{eqn:action}
\end{equation}
where we use the Einstein summation convention over repeated indices. Greek indices describe spacetime, going from $0,\dots,3$, upper-case Latin indices describe the number of fields, going from $1,\dots,N_f$, and lower-case Latin indices describe space, going from $1, \dots, 3$.  The field space indices are raised using the Kronecker delta $\delta^{IJ}$.  The determinant of the spatial metric $g_{\mu \nu}$ is $g$.  In this paper we only consider inflation models with minimal coupling to Einstein gravity and a matter sector described by scalar fields.  The current incarnation of \MMC only solves models with canonical kinetic terms, but we give the equations of motion for models with a non-trivial field-space metric in Appendix~\ref{app:fieldmetric}.  Implementing these general field-space metrics is straightforward since \MMC has been written modularly, but is left for future work.

First-order, non-interacting perturbation theory separates the homogeneous, classical background from the spatially-dependent modes as $\phi_I(t, \vec{x}) \to \phi_I(t) + \delta \phi_I(t, \vec{x})$,
where we assume that these two components can be treated independently.
The homogeneous background fields obey the Klein--Gordon equations
\begin{equation}
\ddot{\phi}_I + 3H \dot{\phi}_I + \frac{\pd V}{\pd \phi^I} = 0   ,
\label{eqn:KGtime}
\end{equation}
where an overdot indicates a derivative with respect to cosmic time $t$ and we use $\mpl^2 = (8 \pi G)^{-1} =1 $ throughout this paper.  The 0-0 Einstein field equation gives the Friedmann equation
\begin{equation}
  3H^2 = \frac{1}{2} \dot{\phi}_I\dot{\phi}^I\ + V(\phi_I)  ,
  \label{eqn:fried}
\end{equation}
which can be differentiated with respect to $t$ to yield
\begin{equation}
2\dot{H} = - \dot \phi_0^2  .
\label{eqn:clock}
\end{equation}
In Eq.~\eqref{eqn:clock} we have used the inflaton \emph{trajectory velocity}, $\dot \phi_0^2 \equiv \dot{\phi}_I \dot \phi^I$.  We can regard the composite field $\phi_0$ as the clock of multifield inflation. It is the classical field defined along the inflaton trajectory, and represents the length of the classical field-space path.

In practice, if the dynamics are inflationary, it is numerically convenient to evolve the equation with the number of $e$-folds $N_e \equiv \ln a(t)$ as the independent variable, giving
\begin{equation}
\frac{\ud^2 \phi_I}{\ud N_e^2 } + (3 -\epsilon) \frac{\ud \phi_I}{\ud N_e} + \frac{1}{H^2}\frac{\pd V}{\pd \phi^I} = 0  ,
\label{eqn:back}
\end{equation}
where we have defined the first slow-roll parameter as
\begin{equation}
  \epsilon\equiv-\frac{\dot H}{H^2}
  =\frac{1}{2} \frac{\ud \phi_I}{\ud N_e} \frac{\ud \phi^I}{\ud N_e}.
  \label{eqn:eps}
\end{equation}
The Friedmann equation~\eqref{eqn:fried} can then also be expressed as
\begin{equation}
H^2 = \frac{V}{3 - \epsilon} .
\label{eqn:H_eps}
\end{equation}
If $V \approx 0$, Eq.~\eqref{eqn:H_eps} requires $\epsilon \approx 3$, which will result in numerical instability whenever we try to set initial conditions that are dominated by their kinetic energy.  We side-step this issue by using the cosmic time Eq.~\eqref{eqn:KGtime} and $H$ as defined in Eq.~\eqref{eqn:fried}.

Solving Eq.~\eqref{eqn:back} therefore only requires the initial conditions $\phi_I$ and $\ud \phi_I/\ud N_e$, because the dependence on the scale factor $a$ is explicitly removed by the 0-0 Einstein equation~\eqref{eqn:H_eps} as a result of assuming a flat FLRW spacetime.  As mentioned in \S\ref{ssect:highlight}, the perturbation spectrum depends on these initial conditions, which are specified as a prior probability distribution $P(\phi_I, \phi_I^{\prime})$.

\subsection{Mode equations}
\label{ssect:mode}

To obtain the first-order equation of motion for the perturbations $\delta \phi_I$, we need to expand the action~\eqref{eqn:action} to second-order and include the first-order scalar perturbations to the flat FLRW metric, given by
\begin{equation}
  \ud s^2 = -  \left(1 + 2 \Phi \right) \ud t^2 - 2 \, a^2 B_{,i} \, \ud t \,  \ud x^i + a^2 \left[ \left(1 - 2\Psi \right) \delta_{ij} - 2 \partial_{\langle i} \partial_{j \rangle} E \right] \ud x^i \ud x^j  ,
  \label{eqn:pertFRW}
\end{equation}
where
\begin{equation}
  \partial_{\langle i} \partial_{j \rangle} E \equiv  \partial_{ i} \partial_{j } E - \frac{1}{3} \delta_{ij} \nabla^2 E
  \label{eqn:XXX}
\end{equation}
 is trace-free.
We choose the spatially-flat gauge, so that $\Psi = E =0$, and vary the expanded action $\delta S_\phi$ with respect to the perturbations $\delta \phi_I(t,\vec x)$ to get the first-order equation of motion for the free-field perturbations.
After Fourier-transforming the scalar perturbations to $\delta \phi_I(\bk)$, the mode equations in this gauge are
\begin{equation}
\frac{\ud^2 \delta \phi_I}{\ud N_e^2} + (3-\epsilon) \frac{\ud \delta \phi_I}{\ud N_e} + \frac{k^2}{a^2H^2} \delta \phi_I +  C_{IJ} \delta \phi^J = 0,
\label{eqn:dphi_mode}
\end{equation}
where
\begin{equation}\label{cij}
C_{IJ} \equiv \frac{\pd_I\pd_J V}{H^2} + \frac{1}{H^2} \left(\frac{\ud \phi_I}{\ud N_e} \pd_J V  +  \frac{\ud \phi_J}{\ud N_e} \pd_I V \right)
+ (3-\epsilon) \frac{\ud \phi_I}{\ud N_e} \frac{\ud \phi_J}{\ud N_e}
\end{equation}
and $\pd_I \equiv \pd/\pd \phi_I$.
The equation of motion for the tensor metric perturbations can be derived similarly; since the non-gauge degrees of freedom are massless and only minimally coupled to the matter sector, the resulting equations of motion are identical to the case of single-field inflation.

To solve the perturbation equations, it is usually convenient to work with the Mukhanov--Sasaki variable $u_I \equiv a \delta \phi_I$.  The mode equation for $u_I$ is
\begin{equation}
\frac{\ud^2 u_I}{\ud N_e^2} + (1-\epsilon) \frac{\ud u_I}{\ud N_e} + \left(\frac{k^2}{a^2H^2} - 2 + \epsilon \right) u_I + C_{IJ} u^J = 0
\label{eqn:ptbmodeeqn}
\end{equation}
with $C_{IJ}$ as in Eq.~\eqref{cij}.
Since the mass matrix, defined as $m^2_{IJ} \equiv \pd_I\pd_J V$, is not necessarily diagonal, the perturbation equations~\eqref{eqn:ptbmodeeqn} mix the annihilation operators for all of the fields \cite{Salopek:1988qh}. We therefore need to expand each perturbation mode $u_I(\bk)$ and $u_I^\dagger(\bk)$ using $N_f$ harmonic oscillators $a_J(\bk)$:
\begin{equation}
  u_I(\bk, N_e) = \psi_{I}^{\;\; J} (\bk, N_e) a_J(\bk) \qquad \mathrm{and} \qquad
u_I^\dagger(\bk, N_e) = \psi_{I}^{\;\;J, *}(\bk, N_e) a_J^\dagger(\bk) ,
\end{equation}
where $(\dagger)$ and $(*)$ represent Hermitian and complex conjugation, respectively.\footnote{An alternative approach is to simply bypass this issue by solving for the field correlation functions directly rather than the individual modes, as in the transport method \cite{Mulryne:2009kh, Mulryne:2010rp, Seery:2012vj, Mulryne:2013uka}. }
We can then define canonical commutation relations $ [ a_J(\bk), a^\dagger_I(\bk')] = (2\pi)^3 \delta_{IJ} \delta^{(3)}(\bk - \bk')$. The mode matrix $\psi_{IJ}$ evolves according to
\begin{equation}
\frac{\ud^2 \psi_{IJ}}{\ud N_e^2} + (1-\epsilon) \frac{\ud \psi_{IJ}}{\ud N_e} + \left(\frac{k^2}{a^2H^2} - 2 + \epsilon \right) \psi_{IJ} + C_{IL} \psi^{L}_{\; \, J} = 0 .
\label{eqn:psi}
\end{equation}
Finding the perturbation spectrum requires setting initial conditions in Eq.~\eqref{eqn:psi} and using the background equations~\eqref{eqn:back} to find the time $N_{e,\mathbf{k}}$ when the mode $\mathbf k$ leaves the horizon, which also depends on the moment at which the pivot scale $k_*$ leaves the horizon, $N_*$ $e$-folds before the end of inflation.

The usual initial condition is the Bunch-Davies state~\cite{Bunch:1978yq}, which assumes the field basis has been chosen such that the $\psi_{IJ}$ are originally diagonal and sets the initial condition for Eq.~\eqref{eqn:psi} as if the mode matrix were freely oscillating in Minkowski space.  This is well-motivated, since for modes deep in the horizon $k \gg aH$, the mode matrix $\psi_{IJ}$ obeys the free wave equation in conformal time
\begin{equation}
\frac{\ud^2 \psi_{IJ}}{\ud \tau^2} + k^2 \psi_{IJ} = 0,
\label{eqn:psi_conf}
\end{equation}
where $d\tau \equiv a \, dt$.
If we assume that the mode matrix is initially diagonal at $\tau = -\infty$, then Eq.~\eqref{eqn:psi_conf} yields two solutions
\begin{eqnarray}
\psi_{IJ} = \frac{1}{\sqrt{2k}}\left( C_1 e^{ik\tau} + C_2 e^{-ik\tau} \right) \delta_{IJ}.
\end{eqnarray}
Translating to $e$-fold time, the initial conditions can be set by
\begin{equation}
  \label{eqn:BD}
 \psi_{IJ}\Big|_{N_e=0} = \frac{1}{\sqrt{2k}}\, (C_1 + C_2)\, \delta_{IJ}
 \qquad \mathrm{and} \qquad
\frac{\ud \psi_{IJ}}{\ud N_e}\Big|_{N_e=0} = \frac{i}{aH}\sqrt{\frac{k}{2}}\,(C_1 - C_2)\,\delta_{IJ} ~.
\end{equation}
The Bunch-Davies initial condition is equivalent to choosing $C_1 = 0$ and $C_2 = 1$.  While only the Bunch-Davies initial condition is implemented in \MMC, non--Bunch-Davies modes could be easily accommodated.\footnote{One would do this by changing the modes' initial conditions in the \code{set\_background\_and\_mode\_ic()} subroutine in \code{modpk.f90}.}

Although the $u_I$'s are convenient for short wavelength modes, they grow exponentially after the modes exit the horizon. So once the mode is outside the horizon, \MMC switches from $u_I$ to $\delta \phi_I$ by matching boundary conditions at a time $N_e^*$ just after horizon crossing with
\begin{equation}
  u_I\Big|_{N_e^*} = e^{N_e^*} \delta \phi_I \Big|_{N_e^*}  \qquad \mathrm{and} \qquad
\frac{\ud u_I}{\ud N_e}\Big|_{N_e^*} = e^{N_e^*} \left(\delta \phi_I + \frac{\ud \delta \phi_I}{\ud N_e}\right)\Big|_{N_e^*}.
\label{eqn:changevar}
\end{equation}

\subsection{Power spectra}

Unlike single-field inflation, the multifield power spectrum involves contractions of the mode matrix.  Using the canonical commutation relations above, the two-point VEV of the field perturbations yields the power spectrum
\begin{equation}
  P_{\delta \phi}^{IJ} (k) = \frac{k^3}{2\pi^2} \left[\frac{1}{a^2}\right] \psi^{I}_{\; L} \; \psi^{J L,*} \, .
  \label{eqn:power}
\end{equation}
When the field trajectories are not turning, on super-horizon scales the fields $\phi_I$ and their momenta $\pi_I$ commute, indicating that they have transitioned to a regime where Eq.~\eqref{eqn:power} can be interpreted as an expectation value over realizations of classical, random fields.

To relate this field-space power spectrum to gauge-invariant perturbation variables \cite{Bardeen:1980kt,PhysRevD.28.679,GrootNibbelink:2000vx}, we first define the curvature perturbation on comoving hypersurfaces $\calR$ by
\begin{equation}
  \calR \equiv \Psi + \frac{1}{3} \nabla^2 E + aH \left( B + v \right) ,
  \label{eqn:rginvariant}
\end{equation}
where $v$ is given in terms of the momentum density of the stress-energy tensor $T^{\mu}_{\; \; \nu}$ as
\begin{equation}
  T^{i}_{\; \, 0} \equiv \left( \bar \rho + \bar P \right) \delta^{ij} \frac{\pd v}{\pd x^j}  ,
  \label{eqn:XXX}
\end{equation}
where $\bar \rho$ and $\bar P$ are the background energy and pressure densities, respectively.
If we evaluate Eq.~\eqref{eqn:rginvariant} on spatially-flat hypersurfaces during inflation, $\calR$ reduces to
\begin{equation}
  \calR = -\frac{H}{\dot \phi_0} \; \omega_I \delta \phi^I  ,
  \label{eqn:rptb}
\end{equation}
where $\omega_I \equiv \dot \phi_I/ \dot \phi_0$ is a basis vector that projects $\delta \phi_I$ along the direction of the classical background trajectory, given by the solutions to Eq.~\eqref{eqn:back}.  The vector $\vec \omega$ and a complementary set of $(N_f-1)$ mutually orthonormal basis vectors $\vec s_K$ form the kinematic basis~\cite{Gordon:2000hv,GrootNibbelink:2001qt}, where the separation between the adiabatic perturbations in Eq.~\eqref{eqn:rptb} and transverse, isocurvature perturbations is made explicit.  Since $\vec \omega$ depends on the nonlinear background evolution, in \MMC we find the $\vec s_K$ numerically by Gram--Schmidt orthogonalization.

The \emph{adiabatic curvature power spectrum} $\PR$ is then the projection of $\PIJ$ along the field vector $\omega_I$, scaled by the pre-factor in Eq.~\eqref{eqn:rptb}, leaving
\begin{equation}
  \PR (k) = \frac{1}{2 \epsilon} \omega_I  \omega_J \PIJ(k)  .
  \label{eqn:pad}
\end{equation}
The gauge-invariant scalar density spectrum in Eq.~\eqref{eqn:pad} is the final result for the adiabatic two-point function to first-order in perturbation theory.

Since Eqs.~\eqref{eqn:rptb}~and~\eqref{eqn:pad} are projected along $\vec \omega$, a simple definition for the isocurvature perturbations $\calS_K$ is the orthogonal projection along the $\vec s_K$ directions
\begin{equation}
  \calS_K \equiv - \frac{H}{\df}  s_{K}^{\, \; J} \delta \phi_J  .
  \label{eqn:iso}
\end{equation}
By projecting $\PIJ$ onto all the directions $s_K$ that are orthogonal to $\omega_I$ and scaling the result as in Eq.~\eqref{eqn:pad}, leads to the \emph{isocurvature power spectrum}:
\begin{equation}
  \PS (k) = \frac{1}{2 \epsilon}\sum_{KL}^{N_f -1} \, \sum_{IJ}^{N_f} s_I^{\; \, K} s_J^{\; \, L} \, \PIJ(k),
  \label{eqn:piso}
\end{equation}
where we have left the summations explicit to indicate that the isocurvature basis vectors are $(N_f-1)$--dimensional.  We include this definition of isocurvature because it is numerically stable, as we discuss in Sect.~\ref{ssect:iso}.

Similarly, we define the \emph{adiabatic-isocurvature cross-spectra} $\Cross$, which is the cross-correlation between the comoving curvature perturbation and the total isocurvature perturbation, given by the contraction of $\PIJ$ with both $\omega$ and the isocurvature basis vectors $s_K$
\begin{equation}
  \Cross(k) = \frac{1}{2 \epsilon} \sum_{K}^{N_f-1} \, \sum_{IJ}^{N_f} \omega_I s_K^J \left( \PIJ +{\mathcal{P}_{\delta \phi}^{JI}}  \right)  .
  \label{eqn:cross}
\end{equation}
Cross-correlations are generically expected if the background trajectory is curved as modes of interest leave the horizon.  By parametrizing Eq.~\eqref{eqn:cross} with the scalar value
\begin{equation}
  \cos \Delta \equiv \frac{\Cross}{\sqrt{\PR \, \PS}},
  \label{eqn:Delta}
\end{equation}
it was shown in Ref.~\cite{Byrnes:2006fr} that, for the case of $N_f = 2$, the value of $r$ is suppressed relative to the single-field, slow-roll expectation by $r \approx 16 \epsilon \sin^2 \Delta,$ to first-order in slow-roll.  In principle, $\Delta$ may be detectable from CMB observations~\cite{Bartolo:2001rt,Wands:2002bn}.

However, by differentiating Eq.~\eqref{eqn:rginvariant} with respect to time $t$, the comoving curvature perturbation will not necessarily be constant even for $k \ll aH$.  Instead,
\begin{equation}
  \dot \calR = -\frac{H}{\df^2} \dPnad  ,
  \label{eqn:XXX}
\end{equation}
where $\dPnad$ is the non-adiabatic pressure perturbation~\cite{GarciaBellido:1995qq,Wands:2000dp,Malik:2002jb}.  This quantity is the difference between the total pressure perturbation
\begin{equation}
  \delta P =  \dot \phi_I \dot{\delta \phi^I} - \dot{\phi}_I\dot{\phi}^I \Phi - V_{,I} \delta \phi^I   ,
  \label{eqn:pptb}
\end{equation}
and the adiabatic pressure perturbation $\delta P_\mathrm{ad} = c_s^2 \delta \rho$, where the speed of sound is $c_s^2 = \dot  P / \dot  \rho$ and the lapse function is
\begin{equation}
  \Phi = \frac{1}{2H} \dot \phi_I \delta \phi^I,
  \label{eqn:lapse}
\end{equation}
defined in the spatially-flat gauge~\cite{Malik:2001rm}.  Given the total density perturbation
\begin{equation}
  \delta \rho =\dot \phi_I \dot{\delta \phi^I} - \dot{ \phi}_I\dot{ \phi}^I \Phi +V_{,I} \delta \phi^I  ,
  \label{eqn:rhoptb}
\end{equation}
the \emph{non-adiabatic pressure power spectrum} $\Ppnad$ reduces to
\begin{eqnarray}
  \Ppnad (k) = \frac{k^3}{2 \pi^2 a^2}
    & \left[
A^I A^J \psi_{I}^{\; \, L} \psi_{LJ}^* + A^I B^J \psi_{I}^{\; \, L} \psi_{LJ}^\prime \right. \\
    &\left. + B^I A^J \psi_{J}^{* \; \,L} \psi_{LI}^\prime + B^I B^J \psi_{I}^{\prime \; \, L} \psi_{LJ}^{* \prime}
  \right] , \notag
  \label{eqn:ppnad}
\end{eqnarray}
where $(\prime)$ indicates a derivative with respect to $e$-foldings $N_e$ and we have defined the vectors
\begin{equation}
  \label{eqn:avect}
  A_I =  \frac{1}{3 a H^2 \epsilon} \phi^{\prime,L}
  \left[
    \left( -3 H^2 \phi_L^\prime - \partial_L V \right) \partial_I V
    + H^2 \partial_M V \, \phi^{\prime, M} \left( \delta_{LI} + \frac{1}{2} \phi_L^\prime \phi_I^\prime \right)
  \right]
\end{equation}
and
\begin{equation}
  B_I = \left(1- c_s^2 \right) H^2 \phi_I^\prime.
  \label{eqn:bvect}
\end{equation}

By analogy to Eq.~\eqref{eqn:iso}, we can build an entropy perturbation from the non-adiabatic pressure perturbation \cite{Gordon:2000hv,Malik:2004tf,Huston:2011fr}, with
\begin{equation}
  \delta S = \frac{H}{\dot P} \dPnad.
  \label{eqn:entptb}
\end{equation}
From this we obtain our final definition of isocurvature, the \emph{comoving entropy spectrum}, given by
\begin{equation}
  \Pent (k) = \left( \frac{H}{\dot P} \right)^2 \Ppnad .
  \label{eqn:pent}
\end{equation}

\subsection{$\delta N$ formalism}
\label{sect:dN}
The separate-universe assumption \cite{Starobinsky:1986fxa,Lyth:1984gv,Sasaki:1995aw,Salopek:1990jq,Sasaki:1998ug,Wands:2000dp}, often referred to as $\delta N$, states that when smoothed on some physical scale much larger than the horizon, the evolution of each smoothed patch can be computed using only background quantities. By identifying that $\zeta = \delta N$, where $\zeta$ is the curvature perturbation on constant density hypersurfaces and $\delta N$ measures the variation in the number of $e$-folds between an initial flat hypersurface and a subsequent constant density hypersurface, Lyth and Rodriguez demonstrated that this assumption can be taken advantage of when computing correlation functions by performing a Taylor expansion in terms of the initial conditions  \cite{Lyth:2005fi}.
\begin{equation}
\zeta =N_{,I} \delta \phi^{I}_{*} + \frac{1}{2}N_{,IJ}\delta \phi_{*}^I \delta \phi_{*}^J + \dots\, .
  \label{eqn:deltaN}
\end{equation}
The main difficulty in this approach lies in computing the derivatives of the number of $e$-folds ($N_{,I} \equiv \partial N_e / \partial \phi_{I,*}$, $N,_{IJ}$ \emph{etc.}). However for sum-separable models these expressions can be computed analytically~\cite{Vernizzi:2006ve,Battefeld:2006sz}. For models with fields much lighter than $H$ at horizon crossing, the numerically intensive calculation of solving for the modes may therefore be unnecessary. \MMC  implements this $\delta N$ slow-roll formalism where we assume that $t_*$ is the moment when the pivot-scale $k_*$ leaves the horizon and that the field perturbations at this time are uncorrelated, with a power spectrum
\begin{equation}
  \mathcal P_{\delta \phi}^{IJ} = \left( \frac{H}{2 \pi} \right)^2 \delta^{IJ}.
  \label{eqn:XXX}
\end{equation}
We also assume that the tensor modes, which are massless and uncoupled to the matter sector, have a power spectrum $\mathcal P_{h} = 8 \, (H/2 \pi)^2$.
At least to first order, on super-horizon scales $\zeta = \mathcal R$ \cite{Malik:2008im}, which allows us to compare the predicted power spectrum for $\zeta$ using the $\delta N$ formalism to the adiabatic power spectrum in Eq.~\eqref{eqn:pad}.

If the potential $V$ is sum-separable so that
\begin{equation}
  V = \sum_{I} V_I(\phi_I),
  \label{eqn:XXX}
\end{equation}
then we can use the Klein--Gordon equations~\eqref{eqn:back} for the scalar fields to obtain a sum-separable expression for the amount of expansion between the two surfaces
\begin{equation}
  N_e = -  \sum_I \int_*^c \frac{V_I}{V_I'} d \phi_I,
  \label{eqn:N}
\end{equation}
where $V_I' \equiv dV_I/d\phi_I$.  If $V$ were not sum-separable, the derivatives of $N_e$ would in general have to be obtained numerically by evolving the background equations of motion~\eqref{eqn:back} on a stencil in field-space and taking the finite difference.  We have not implemented this feature in \MMC as it is more computationally intensive than solving the mode equations.

When the potential is sum-separable, the derivatives of $N_e$ can be simplified into the expressions~\cite{Vernizzi:2006ve,Battefeld:2006sz}
\begin{equation}
  N_{,I} = \frac{1}{\sqrt{2 \epsilon_{I}^*}} \frac{V_I^* + Z_I^c}{V^*}
  \label{eqn:XXX}
\end{equation}
and
\begin{equation}
 N_{,IJ} = \delta_{IJ} \left[1 - \frac{\eta_I^*}{2 \epsilon_I^*} \left( \frac{V_I^* + Z_I^c}{V^*} \right) \right]
  + \frac{1}{\sqrt{2 \epsilon_J^*} V^*} \frac{ \partial Z_J^c}{\partial \phi_I^*},
  \label{eqn:XXX}
\end{equation}
where
\begin{equation}
  Z_I^c = V^c \frac{\epsilon_I^c}{\epsilon^c} - V_I^c,
  \label{eqn:XXX}
\end{equation}
\begin{equation}
  Z_{IJ}^c = - \frac{V_c^2}{V^*} \sqrt{\frac{2}{\epsilon_J}}
  \left[\sum_{K=1}^{N_f} \epsilon_K \left( \frac{\epsilon_I}{\epsilon} - \delta_{IK} \right) \left( \frac{\epsilon_J}{\epsilon} - \delta_{JK}\right) \left( 1 - \frac{\eta_K}{\epsilon} \right) \right]_c,
  \label{eqn:XXX}
\end{equation}
and the slow-roll parameters are
\begin{equation}
  \epsilon \equiv \sum_I \epsilon_I = \frac{1}{2} \sum_I \frac{V_I'^2}{V^2}
  \label{eqn:XXX}
\end{equation}
and
\begin{equation}
      \eta \equiv \sum_I \eta_I = \sum_I \frac{V_I''}{V}.
  \label{eqn:XXX}
\end{equation}
The contribution from the EOI surface is therefore completely encoded in the functions $Z_I$ and its derivatives.

The relationship~\eqref{eqn:deltaN} and the expansion equation~\eqref{eqn:N} allow us to define pivot-scale observables for the scalar perturbations $\zeta$.  We will focus on the observables obtainable only through the first and second derivatives of $N_e$, and express our results only to the lowest order in slow-roll.  We start with the $\zeta$ power spectrum
\begin{equation}
  \mathcal P_\zeta = N_{,I} N^{,I} \left( \frac{H}{2 \pi} \right)^2,
  \label{eqn:XXX}
\end{equation}
and the tensor-to-scalar ratio
\begin{equation}
  r = \frac{8}{N_{,I} N^{,I}},
  \label{eqn:XXX}
\end{equation}
which have simple expressions only in terms of $N_{,I}$.  The adiabatic and tensor spectral indices $n_s$ and $n_t$ also have easily evaluated expressions
\begin{equation}
  n_s -1 = -2 \epsilon_* - \frac{2}{N_{,I} N^{,I}} +
  \left( \frac{2}{V} \right) \frac{V_{,IJ} N^{,I} N^{,J}}{ N_{,K} N^{,K}}
  \label{eqn:XXX}
\end{equation}
and
\begin{equation}
  n_t = \frac{ -2 \epsilon_* } {1- \epsilon_*} .
  \label{eqn:XXX}
\end{equation}
The expression for the scalar running $\alpha_s$ is more complicated, but straightforward to compute (\emph{e.g.}, Eq. 6.14 in Ref.~\cite{Dias:2012nf}).

To obtain the amplitude of the predicted non-Gaussianity we further assume that the field perturbations at horizon crossing are purely Gaussian, since the non-Gaussianity generated by sub-horizon evolution of the modes is typically slow-roll suppressed~\cite{Seery:2005gb,Vernizzi:2006ve}, assuming that slow-roll is not violated.  Following Refs.~\cite{Maldacena:2002vr,Vernizzi:2006ve}, we use the non-linearity parameter
\begin{equation}
  - \frac{6}{5} \fnl \equiv \left[ \frac{\prod_i k_i^3}{\sum_i k_i^3} \right] \frac{B_\zeta}{4 \pi^4 \mathcal P_\zeta^2} \approx \frac{ N_{,I} N_{,J} N^{,IJ}}{ \left( N_{,K} N^{,K} \right)^2},
  \label{eqn:XXX}
\end{equation}
where $B_\zeta$ is the bispectrum.  Given Gaussian field perturbations at horizon crossing, the trispectrum amplitude is then parametrized by the non-linearity parameters \cite{Alabidi:2005qi,Byrnes:2006vq}
\begin{equation}
  \taunl = \frac{ N_{,IJ} N^{,IK} N^{,J} N_{,K}}{ \left( N_{,L} N^{,L} \right)^3}
  \label{eqn:XXX}
\end{equation}
and
\begin{equation}
  g_\mathrm{NL} = \left( \frac{25}{54} \right) \frac{N_{,IJK} N^{,I}N^{,J}N^{,K}}{\left(N_{,L}N^{,L} \right)^3}.
  \label{eqn:XXX}
\end{equation}
Since $g_\mathrm{NL} \sim N_{,IJK}$ we do not compute it here, although it could be implemented by taking the third derivative of $N_e$ as in Ref.~\cite{Battefeld:2006sz}.

\subsection{Bundle width}
\label{ssect:bundle}

An alternative method of monitoring isocurvature is to acknowledge that under slow-roll, the separate universe assumption is precisely analogous to geometrical optics in field space~\cite{Seery:2012vj}. The smoothed spatial patches described in Sect.~\ref{sect:dN} each correspond to a distinct non-interacting trajectory in field space with perturbed initial conditions with respect to some arbitrary fiducial trajectory. These perturbed trajectories can then be thought of as forming a \emph{bundle} moving through a medium with refractive index $\sqrt{2\epsilon}$. One can therefore track isocurvature evolution using only background quantities, by associating isocurvature growth and decay with dilation and contraction of the bundle.  While the precise analogy with geometrical optics does not remain when slow-roll is violated, one still has a useful set of geometrical quantities for understanding the evolution of field perturbations.

Under slow-roll, the Klein--Gordon equation reduces to
\begin{equation}\label{eq:KGSR}
\frac{\ud \phi_{I}}{\ud N_e}=-\partial_{I}\log V,
\end{equation}
which is \emph{Huygen's equation} and an infinitesimal vector propagated along the beam is called a \emph{Jacobi field}. If we take $\delta\phi_{I}$ to be such a field, we can obtain from Eq.~\eqref{eq:KGSR} how it will propagate:
\begin{equation}
\frac{\ud \delta\phi_{I}}{\ud N_e}=-\left [\partial_{I}\partial_{J}\log V\right ] \delta\phi^{J},
\end{equation}
which is the slow-roll analogue of Eq.~\eqref{eqn:ptbmodeeqn} \cite{Seery:2012vj, Mulryne:2013uka}. Indeed, we could have recast the whole of Sect.~\ref{ssect:mode} in this language \cite{Mulryne:2013uka}. The term in square brackets is usually referred to as the expansion tensor and it encodes all information required for tracking field perturbations;\footnote{This point is heavily emphasised in the context of the \emph{transport method} of computing inflationary correlation functions \cite{Mulryne:2009kh, Mulryne:2010rp, Seery:2012vj, Mulryne:2013uka, DiasUnpub}.} under slow-roll it has a particularly simple geometric interpretation. We can decompose the expansion tensor as
\begin{equation}
\partial_{I}\partial_{J}\log V=\frac{\theta}{N_{f}}+\sigma_{IJ}+\omega_{IJ},
\end{equation}
where $\sigma_{IJ}$ is the symmetric shear, $\omega_{IJ}$ is the antisymmetric twist, and the key quantity for our purposes is the dilation, given by the trace
\begin{equation}
\theta= \mathrm{Tr}\, \partial_{I}\partial_{J}\log V.
  \label{eqn:bundle}
\end{equation}
If $\theta>0$, then isocurvature is growing and if $\theta<0$, then isocurvature is decaying. We can then find a measure $\Theta$ of the bundle width  by integrating this along the inflationary trajectory
\begin{equation}\label{eqn:Theta}
\Theta\equiv \exp \left[ \int_{N_{0}}^{N} \theta(N') dN' \right],
\end{equation}
which normalizes the bundle width so that $\Theta (N_0) \equiv 1$.
In situations where we only want to solve the background equations of motion, the bundle width is informative for understanding whether or not $\zeta$ becomes conserved on superhorizon scales, which is a crucial requirement when comparing the predictions of a model with observation.  For two fields $\Theta\rightarrow0$ is a necessary and sufficient condition for the approach to an adiabatic limit.  However when there are more fields the situation is more complicated, \emph{e.g.}, the bundle may contract to a sheet rather than a caustic. We refer the reader to Ref.~\cite{Seery:2012vj, Dias:2012nf, DiasUnpub} for more details.

\section{The method}
\label{sect:method}

We outline the procedure used to obtain the power spectrum predictions, with the algorithmic structure of \MMC in Algorithm~\ref{alg:mmc}.  While this largely follows previous  implementations, such as \textsc{Pyflation} \cite{Huston:2009ac,Huston:2011fr,Huston:2011vt,Huston:2013kgl}, we give the method  the sake of clarity and reproducibility.

\begin{algorithm}
  \caption{\MMC method}
  \label{alg:mmc}
  \begin{algorithmic}
    \State \textbf{define} sample size, $V$, $k_*$
    \ForAll{elements in sample}
      \Procedure{Background Solver:}{}
        \State \textbf{get} Lagrangian parameters for $V$ and ICs for Eq.~\eqref{eqn:back} from prior PDF
        \State \textbf{with} the end-of-inflation (EOI) criterion set by user, \textbf{solve} Eq.~\eqref{eqn:back} until EOI
        \State \textbf{check} inflation ($\ddot a >0$) started and ended
      \EndProcedure

      \Procedure{Scale-factor Normalizer:}{}
        \State \textbf{get} $N_*$ from user \textbf{or} by prior PDF
        \State \textbf{check} total inflationary $e$-folds $N_\mathrm{tot} \ge N_*$
        \State \textbf{define} $a$ such that $k_* = a_* H_*$ at $N_e = N_\mathrm{tot} - N_*$ before inflation ends
      \EndProcedure

      \Procedure{$\delta N$ Calculator:}{}
        \If{$V$ is sum-separable,} calculate $\delta N$ observables near $k_*$
        \EndIf
      \EndProcedure

      \ForAll{modes $k$}

        \Procedure{Mode Initializer:}{}

        \State \textbf{define} initial time $N_{e,0}$ with $k \gg a_0 H_0$
          \State \textbf{while} the corrections to Eq.~\eqref{eqn:psi_conf} are above some tolerance:
          \State \qquad \textbf{set} earlier $N_{e,0}$
          and \textbf{check} $N_{e,0} > 0 $
          \State \textbf{set} Bunch-Davies IC for mode matrix $\psi_{IJ}(\mathbf k)$ at $N_{e,0}$

        \EndProcedure

        \Procedure{Mode Solver:}{}
          \State \textbf{solve} Eq.~\eqref{eqn:ptbmodeeqn} until $k \approx aH$
          \State \textbf{change} variable as in Eq.~\eqref{eqn:changevar} and \textbf{solve} until EOI
          \State \textbf{calculate} power spectra for $k$
        \EndProcedure

      \EndFor

      \Procedure{$k_*$-observable Calculator:}{}
        \State \textbf{calculate} amplitudes, spectral indices, \emph{etc.} at $k_*$ by finite difference in $k$-space
      \EndProcedure

    \EndFor

  \end{algorithmic}
\end{algorithm}

We start by defining the functional form of the potential $V$ and prior probability distribution functions (PDFs) for the parameters that define $V$, which we call Lagrangian parameters or model parameters, and the background initial conditions $\phi_{I,0}$ and $\phi_{I,0}^\prime$.  We treat the simple situation of exactly specifying a set of Lagrangian parameters and initial conditions as a special case, where the prior probability for these parameters is trivial.  Given these priors, the program will build a numerical sample by iteration until a pre-defined number of samples is reached.

\MMC first solves the background equations of motion~\eqref{eqn:back} until the end-of-inflation.  While we have included the natural condition of $\epsilon =1$ as the default ending criterion for inflation, there is complete functionality to end inflation by another method, in particular a waterfall transition via the hybrid mechanism~\cite{Linde:1993cn,Copeland:1994vg} at some reference phase-space point.

Given a value for the number of $e$-folds $N_*$ between when the pivot scale $k_*$ leaves the horizon and the end-of-inflation, which is either fixed by the user or set in each iteration of the code through the sampling of a prior probability $P(N_*)$, we obtain the value of $H$ at horizon crossing by interpolating the numerical background solution.  The pivot scale $k_*$ must be pre-defined by the user and defaults to $0.002 \impc$.  From this, we normalize the size of the universe so that $k_* = a H_*$ at $N_e = N_\mathrm{tot}-N_*$.

For each scale of interest $k$, we  set the modes' initial conditions at a time $N_{e,0}$ when that mode is significantly sub-horizon, $k \gg a_0 H_0$.  For the Bunch-Davis initial state, this point is chosen iteratively by making sure that the relative corrections to Eq.~\eqref{eqn:psi_conf} that are sub-dominant for $k \gg aH$ are smaller than a pre-defined tolerance.  This tolerance is set to $1\e{-5}$; from observing the sub-horizon evolution of the modes, using a tolerance at least this tight gives no change to the value of the modes at horizon crossing.

We then solve the mode equations~\eqref{eqn:ptbmodeeqn} for the variable $\psi_{IJ}$ for the period of time when the modes are smaller than the causal horizon, $k \gtrsim aH$, and then switch to a two-index matrix built from the $u_I$ in Eq.~\eqref{eqn:changevar} for super-horizon evolution.  The power spectra are  calculated for each $k$ and various pivot-scale statistics are evaluated by finite-difference between a few scales $k_i$ near $k_*$.  If the potential $V$ is sum-separable, the program  also calculates the $\delta N$ values for the observables described in Section~\ref{sect:dN}.

Numerous checks are performed on the background and mode equation evolution so that \MMC will either fail gracefully if a fatal exception is raised or declare a particular initial parameter set  invalid and iteratively generate a new set of parameters in order to explore cosmologically relevant parameter sets.  We have extensively tested the numerical stability of the code and have included a number of easily controllable options allowing the user to control the numerical accuracy, as well as the type of ODE solver.

\section{Numerical results}
\label{sect:results}

\subsection{Isocurvature stability}
\label{ssect:iso}

\begin{figure}
  \centering
  \includegraphics{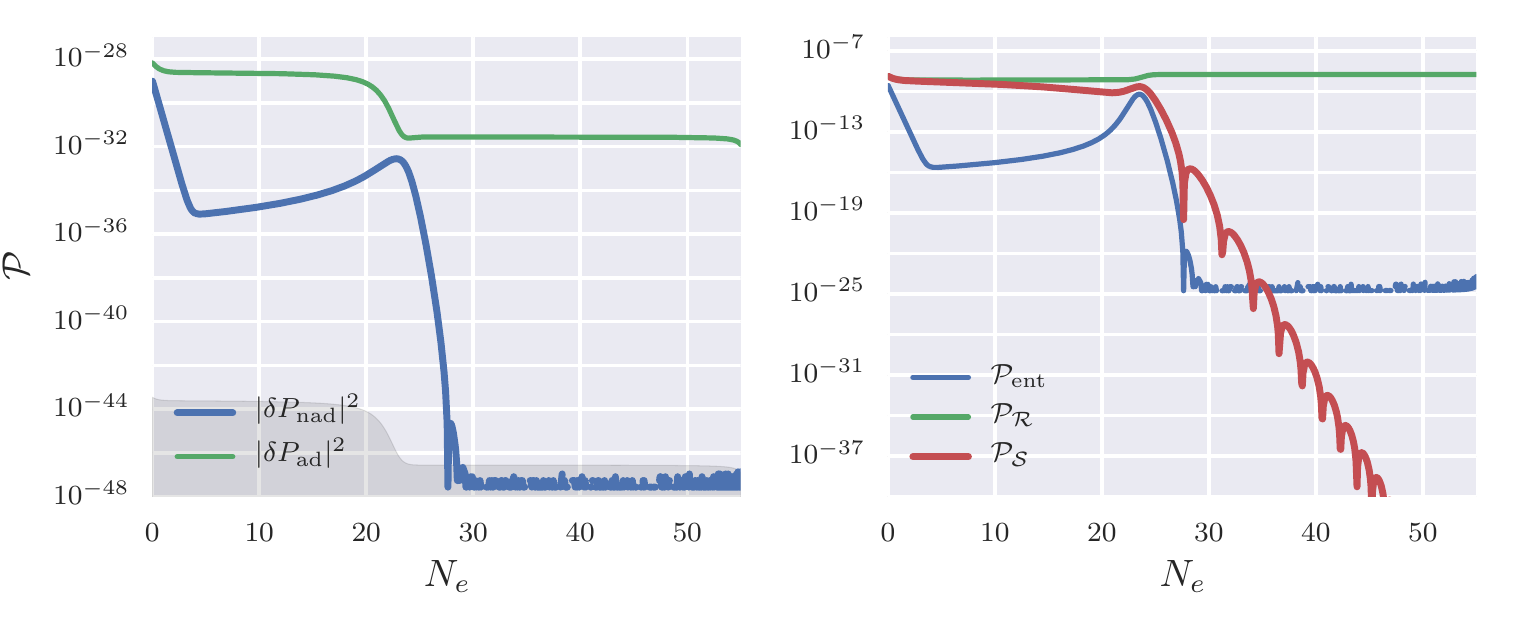}
  \caption{The evolution of the power spectra during the last 55 $e$-folds of inflation for a two-field $N_f$-quadratic model.  (\emph{Left}) The power spectrum for adiabatic (\emph{green}) and non-adiabatic (\emph{blue}) pressure perturbations $\delta P$.
    The total pressure spectrum and the adiabatic pressure spectrum are nearly coincident on this scale, so the total pressure spectrum $\Pp$ has not been plotted.  The gray area is an estimate of the region dominated by double-precision error due to round-off in $\Ppnad$.  (\emph{Right}) The power spectra for perturbations in the adiabatic curvature $\PR$, the isocurvature $\PS$, and the comoving entropy $\Pent$.  $\Pent$ is a rescaling of $\Ppnad$ and is numerically unstable for $N_e \gtrsim 30$ in this model.  $\PS$ is numerically stable until the end of inflation.
  }
  \label{fig:iso}
\end{figure}

Fig.~\ref{fig:iso} illustrates a problem that arises when computing the isocurvature spectra $\Ppnad$ and $\Pent$.  We have plotted the super-horizon evolution of the power spectra for the adiabatic and non-adiabatic pressure perturbations, as well as the adiabatic curvature, entropic, and isocurvature spectra, with $N_*=55$, for a two-field inflation model with the potential
\begin{equation}
  V = \frac{1}{2}m_1^2 \phi_1^2 + \frac{1}{2}m_2^2 \phi_2^2.
  \label{eqn:V_2field}
\end{equation}
To match the analysis performed in Refs.~\cite{Lalak:2007vi,Huston:2011fr,Avgoustidis:2011em,Huston:2013kgl} we choose $m_1^2=10^{-11.7}$, $m_2^2=10^{-10.0}$, and initial conditions $\phi_{1,0} = \phi_{2,0} = 12.0 \, \mpl$.  In particular, Fig.~\ref{fig:iso} can be compared directly to Figs~1~and~3 in Ref.~\cite{Huston:2011fr}.  With this choice of parameters, the background trajectory evolves primarily along the direction of the heavier field $\phi_2$ for $N_e \lesssim 25$, then turns sharply toward the $\phi_1$ direction for the remainder of inflation.  The effect of this turn on the super-horizon perturbations can be seen clearly in the power spectra in Fig.~\ref{fig:iso}.

In general, the calculation of $\Ppnad$ and $\Pent$ becomes dominated by numerical error as the isocurvature perturbations decay.  From Fig.~\ref{fig:iso}, regardless of the amplitude of the isocurvature modes, the adiabatic pressure perturbations $\delta P_\mathrm{ad} = c_s^2 \delta \rho$ do not exponentially decay between horizon exit and the end of inflation.  For the example model~\eqref{eqn:V_2field}, the power spectrum for $\delta P_\mathrm{ad}$ is approximately constant after the turn at $N_e \approx 25$.  However, the total pressure perturbation $\delta P$ is approximately equal to $\delta P_\mathrm{ad}$ during this time and the difference between the two reduces exponentially as the isocurvature modes decay.

Since $\delta P_\mathrm{nad} \equiv \delta P - \delta P_\mathrm{ad}$ and $\delta P_\mathrm{ad} \to \delta P$, the numerical accuracy for $\delta P_\mathrm{nad}$ is limited by the real precision of the computer, which results in a finite difference error in the numerical calculation of $\delta P_\mathrm{nad}$ and a loss of significance.  Using standard double precision accuracy, the expected error in $\delta P_\mathrm{nad}$ should then be
\begin{equation}
  \Delta_\mathrm{err} \Ppnad \sim \mathcal O(10^{-15}) \; \mathcal P_{\delta P} \sim \mathcal O(10^{-15}) \; \mathcal P_{\delta P, \mathrm{ad}},
  \label{eqn:num_error}
\end{equation}
which is confirmed in Fig.~\ref{fig:iso}.  Without correcting for this dominant error term, the value of $\Ppnad$ will oscillate arbitrarily between zero and the limit in Eq.~\eqref{eqn:num_error}, which is an upper bound on the amplitude of the non-adiabatic pressure perturbations.  Since entropic perturbations are usually defined as~\cite{Malik:2004tf}
\begin{equation}
  S_{IJ} \equiv \zeta_I - \zeta_J ,
  \label{eqn:XXX}
\end{equation}
where $\zeta_I$ is the curvature perturbation resulting from the $I^\mathrm{th}$ fluid, this problem will arise naturally for all calculations of $\Pent$.

In contrast, the calculation of $\PS$ in Eq.~\eqref{eqn:piso} is directly proportional to the value of the decaying isocurvature modes in the kinematic basis.  Using this isocurvature spectrum largely alleviates the numerical problems with $\dPnad$, yielding a more faithful measure with a higher degree of accuracy.  Figure~\ref{fig:iso} shows the exponential decay of $\PS$ after the super-horizon turn at $N_e \sim 25$.  We compare this to $\Pent$, which becomes numerically unstable at $N_e \approx 30$, showing that the two measures $\PS$ and $\Pent$ are separated by 27 orders of magnitude at the end of inflation, despite being of the same magnitude at horizon crossing.\footnote{As the adiabatic limit is approached, $\PS$ can also receive a dominant contribution from roundoff error in the Gram-Schmidt orthogonalization procedure.  If some components of the isocurvature vectors $s_K^I$ are much smaller than others, this can result in a spurious projection of $\PR$ onto the isocurvature directions.  In \MMC we have included an optional subroutine \code{renormalize\_remove\_smallest} in \code{modpk\_potential.f90}, where the components of $s_K^I$ are set to zero if they do not affect the normalization of $s_K$, \emph{i.e.}, if the value of $s_K^I$ is indistinguishable from roundoff error.  In practice, we have never seen this problem arise, so this option needs to be uncommented in the source code before compilation.}

\subsection{A case study: $N_f$-flation with a step}
\label{ssect:case}

We have shown in Refs.~\cite{Easther:2013rva,Price:2014ufa} that \MMC is able to produce large volume Monte Carlo samples for $N_f$--monomial inflation with the potential
\begin{equation}
  V = \frac{1}{p} \sum_I \lambda_I |\phi_I|^p,
  \label{eqn:V_monomial}
\end{equation}
for real exponents $p$ \cite{Liddle:1998jc,Kanti:1999vt,Kanti:1999ie,Kaloper:1999gm,Easther:2005zr,Dimopoulos:2005ac,Kim:2006ys,Kim:2006te,Kim:2007bc,Wenren:2014cga}.  In Ref.~\cite{Easther:2013rva} we focused on the $N_f$--quadratic case with $p=2$ and demonstrated that the predictions for the power spectrum do not sensitively depend on the prior probability chosen for the initial conditions of the fields.  In Ref.~\cite{Price:2014ufa} we further demonstrated this for the general case in Eq.~\eqref{eqn:V_monomial}, while focusing on the gravitational wave consistency relation.  We were able to straightforwardly compare the analytical $\delta N$ results to the numerics, greatly simplifying the procedure for comparing analytical results to the full numerical calculation.  We include all of the IC priors used in these papers in \MMC.

Since we have already demonstrated the power of \MMC in Monte Carlo sampling, in this paper we will instead focus on a few case studies that are interesting due to their analytic intractability.
We present results for a multifield generalization of the inflationary step-potential first used in Ref.~\cite{Adams:2001vc}.  This potential has the form
\begin{equation}
  V = \frac{1}{2} \sum_I m_I^2 \phi_I^2 \left[ 1 + c_I \tanh \left( \frac{\phi_I - \bar \Phi_{I}}{d_I} \right) \right]
  \label{eqn:V_step}
\end{equation}
with masses $m_I$ and real constants $d_I$, $c_I$, and $\bar \Phi_I$ specifying the slope, amplitude, and position, respectively, for a step feature in the $I\th$ field.  Phase transitions in sectors coupled only gravitationally to the inflaton sector may generate these hyperbolic-tangent features in $V$ and leave an observable imprint in the primordial density spectra if these symmetry breaking transitions occur during the last $\mathcal O(60)$ $e$-folds of inflation \cite{Adams:1997de,Adams:2001vc}.  In the sharp-step limit, these features introduce oscillations as a function of $k$ into the adiabatic curvature power spectrum and a scale-dependent, oscillatory bispectrum~\cite{Adams:2001vc,Chen:2006xjb,Chen:2008wn,Adshead:2011jq}.  To keep $V>0$ we require $c_I<1$ and to satisfy the latest constraints on oscillations in the scalar power spectrum amplitude requires $c_I \lesssim 10^{-3}$, assuming that the step occurs as the scales relevant for the CMB leave the horizon \cite{Easther:2013kla,Meerburg:2013cla,Meerburg:2013dla}.

With $c_I \to 0$, Eq.~\eqref{eqn:V_step} is an uncoupled assisted inflation model~\cite{Liddle:1998jc,Copeland:1999cs}, first proposed in Ref.~\cite{Dimopoulos:2005ac}.  Models with a step feature are additionally interesting, because they can fit a wider range of data and have been well-studied in the single-field case.  In particular, Ref.~\cite{Adshead:2011jq} contains an elegant analytical calculation for the single-field case of Eq.~\eqref{eqn:V_step}.  However, replicating the same calculation for the general potential would be difficult ---  if not impossible --- with the same techniques, since the possible existence of isocurvature perturbations significantly complicates the analysis.  Consequently, a numerical exploration of this model is well-motivated.

Fixing the number of fields to $N_f=10$, we set the initial conditions to $\phi_{I,0}=10$, with the initial velocities set in slow-roll.  The size and slope of the step are set to $c_I = 10^{-3}$ and $d_I=10^{-2}$ respectively, and the masses $m_I$ relative to the fiducial mass to $\bar m^2 =4.30\e{-11}$, which in the single-field limit yields $A_s$ at the best-fit value from the \emph{Planck} $TT$ data.  Following Ref.~\cite{Easther:2005zr}, we choose the masses $m_I$ according to the Mar\v{c}enko-Pastur distribution
\begin{equation}
  P(m_I^2) = \frac{1}{2 \pi m_I^2 \; \bar m^2  \beta} \; \sqrt{ \left( \beta_+ - m_I^2 \right) \left( m_I^2 - \beta_- \right) },
  \label{eqn:MP}
\end{equation}
where
\begin{equation}
  \beta_{\pm} = \bar m^2 \left( 1 \pm \sqrt{\beta} \right)^2
  \label{eqn:betaeq}
\end{equation}
with $\beta =1/2$.
This distribution of masses is derived in Ref.~\cite{Easther:2005zr}, and has also been used in Refs.~\cite{Kim:2007bc,Battefeld:2008bu,Easther:2013rva,Bachlechner:2014hsa}.

\begin{figure}
  \centering
  \includegraphics{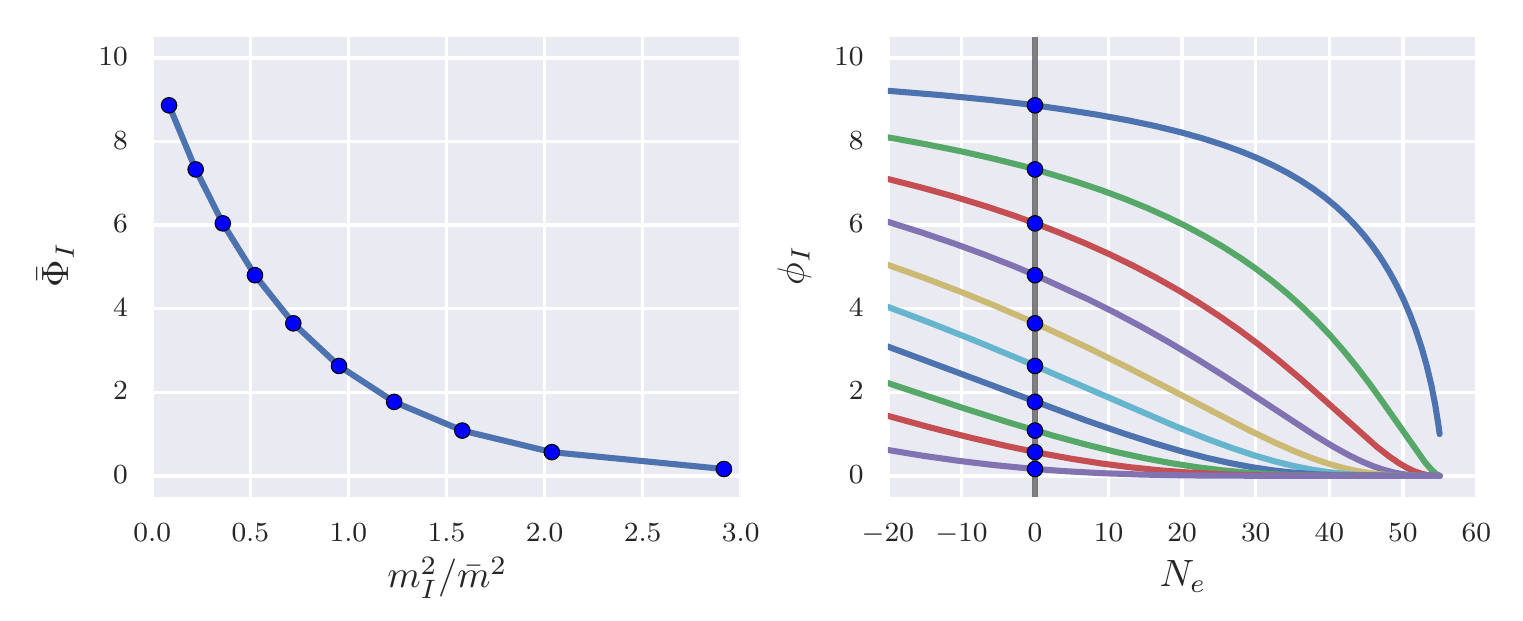}
  \caption{
    (\emph{Left}) The masses $m_I$ for each of the 10 fields in Eq.~\eqref{eqn:V_step}, drawn from the distribution~\eqref{eqn:MP} with $\bar m^2 = 4.3\e{-11}$, compared to the corresponding step positions $\bar \Phi_I$ for that field, which is positioned so that the pivot scale $k_*=0.002\impc$ leaves the horizon at $\bar \Phi_I$, given the initial conditions $\phi_{I,0} = 10$.
    (\emph{Right}) The field trajectories (\emph{colored lines}), with the same initial condition, as a function of $e$-folding $N_e$, with $k_*$ (\emph{vertical line}) leaving the horizon 55 $e$-folds before the end of inflation.  The step positions $\bar \Phi_I$ are marked in blue and $N_e$ has been renormalized so that $k_*=aH$ at $N_e=0$.
  }
  \label{fig:field_back}
\end{figure}

We set the step positions $\bar \Phi_I$ for each field at the field-space point where the pivot scale $k_*=0.002 \impc$ leaves the horizon at $N_*=55$ $e$-folds before the end of inflation in the no-step limit, $c_I \to 0$.
Since the fields have identical initial conditions, the $\bar \Phi_I$ are functions only of the masses, so we plot the step positions versus the $m_I$ in Fig.~\ref{fig:field_back}.  We also present the field-space trajectories according to Eq.~\eqref{eqn:back} for the last 75 $e$-folds of inflation with these parameters.  The heavier fields relax more quickly toward their minimum at $\phi_I =0$ and the lighter fields have a larger value at horizon crossing.  Since $c_I = 10^{-3}$, the step is not obviously visible at the level of the background trajectory without zooming in significantly.

\begin{figure}
  \centering
  \includegraphics{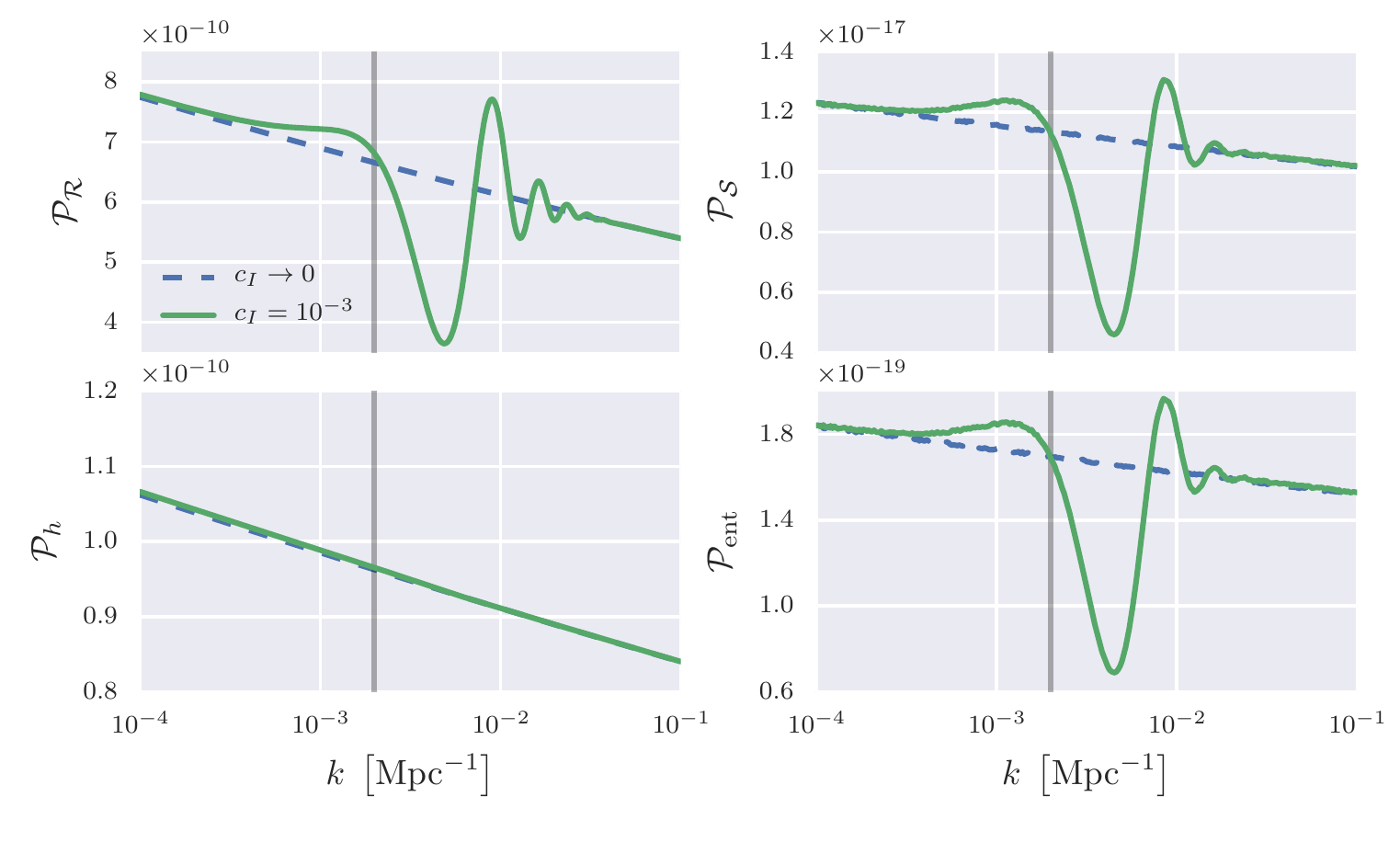}
  \caption{Features in the power spectra due to the step~\eqref{eqn:V_step}, which is positioned so that it affects the power spectra around the pivot scale $k_*= 0.002 \impc$ (\emph{gray}).  We compare (\emph{dashed, blue}) the no-step case with $c_I=0$, to (\emph{solid, green}) the case with $c_I = 10^{-3}$.   While there are oscillations in the adiabatic $\PR$, isocurvature $\PS$, and entropic $\Pent$ spectra, there is little variation in the tensor spectrum $\Ph$.
  }
  \label{fig:pk_compare}
\end{figure}

However, Fig.~\ref{fig:pk_compare} shows the substantial effect on the power spectra due to the steps.  We see oscillatory behavior in the adiabatic, isocurvature, and entropic power spectra, but almost no change in the tensor spectrum. Furthermore, we can see clearly that $\PS$ and $\Pent$ exhibit a nearly identical feature, simply scaled by a factor of roughly 65.  These features in the isocurvature spectrum may lead to interesting effects during reheating or the subsequent evolution of the post-inflation universe.

\section{Conclusion}
\label{sect:conclusion}

We present the Fortran 95/2000 code \MMC, designed to maximize computational efficiency when numerically exploring a broad range of multifield inflation models. The code also provides Monte Carlo sampling of prior probabilities for inflationary model parameters and initial conditions, enabling automated model exploration and the computation of probability distributions for observables.  The mode equation method has a broad range of applicability, but the computational cost scales with the number of fields as $\mathcal O(N_{f}^{2})$.   For models with sum-separable potentials, we have also implemented a slow-roll $\delta N$ calculation, which only requires solving the background equations of motion once in order to obtain the full power spectrum as well as higher order statistics.  This drastically improves computation time, since the background equations of motion are only $\mathcal O(N_f)$.

This code was used to explore the predictions of models with $\order{100}$ fields in Refs.~\cite{Easther:2013rva,Price:2014ufa}; here, we demonstrated its use with an $N_f$-flation model with a step.  We find that a feature in the inflationary potential not only results in a feature in both the adiabatic power spectrum as a function of scale, $\PR(k)$, as well as the isocurvature spectra $\PS$, $\Pent$, and $\Ppnad$, with possible implications for the dynamics of many-field preheating scenarios.  Further, we see numerical evidence that the isocurvature spectrum $\PS$ is a simple rescaling of the entropic spectrum $\Pent$, indicating that the projection of the mode power spectrum onto the isocurvature directions is related to a quantity that sources a change in $\mathcal R$ on super-horizon scales.

\MMC  complements codes that currently exist to numerically compute the inflationary power spectra~\cite{Adams:2001vc, Ringeval:2005yn, Martin:2006rs, Huston:2009ac,Mortonson:2010er, Easther:2011yq, Huston:2011fr,Huston:2011vt,Norena:2012rs,Huston:2013kgl, DiasUnpub}.  We provide a basic usage manual for \MMC in Appendix~\ref{app:usage} to help users to adapt this program to their own problems. The theoretical basis of the method is outlined in Section~\ref{sect:method}. The ability of \MMC to solve numerically challenging problems, such as the step-potential in \S\ref{ssect:case}, and to provide large samples of many-field inflationary models adds significantly to the early universe cosmologist's toolkit for exploring and understanding realistic inflation models.

\acknowledgments

We thank Grigor Aslanyan, Adam Christopherson, Mafalda Dias, Ian Huston, Karim Malik, David Mulryne, David Seery, and Jonathan White for many helpful discussions throughout the duration of this work.  We acknowledge the use of the \code{dvode\_f90\_m.f90} numerical integration code,\footnote{\url{www.radford.edu/~thompson/vodef90web/}} developed by G. Byrne, S. Thompson, and LLNL, which we redistribute here under the terms of their BSD-like license, and the CSV writing capabilities of the FLIBS open-source library,\footnote{\url{http://flibs.sourceforge.net/}} developed by Arjen Markus.
JF is supported by IKERBASQUE, the Basque Foundation for Science. HVP is supported by STFC and the European Research Council under the European Community's Seventh Framework Programme (FP7/2007-2013) / ERC grant agreement no 306478-CosmicDawn. The authors acknowledge the contribution of the NeSI high-performance computing facilities and the staff at the Centre for eResearch at the University of Auckland. New Zealand's national facilities are provided by the New Zealand eScience Infrastructure (NeSI) and funded jointly by NeSI's collaborator institutions and through the Ministry of Business, Innovation \& Employment's Research Infrastructure programme [{\url{http://www.nesi.org.nz}}].  This work has been facilitated by the Royal Society under their International Exchanges Scheme.  This work was supported in part by National Science Foundation Grant No. PHYS-1066293 and the hospitality of the Aspen Center for Physics.

\appendix

\section{Non-canonical kinetic terms}
\label{app:fieldmetric}

This appendix largely follows Ref.~\cite{Sasaki:1995aw} and describes the background and first-order mode equations for inflation models with multiple scalar fields, but a general field space metric $G_{IJ}(\phi_K)$.  These equations are not coded into \MMC, but are an important reference, since multifield models with non-canonical kinetic energy terms fit the \emph{Planck} data extremely well \cite{Kaiser:2013sna}.  These have also been implemented into the single-field version of \textsc{ModeCode} by Ref.~\cite{Li:2012vta}.\footnote{We will update \MMC to give this functionality in the near future and a simple implementation in Mathematica using the Transport method will be available soon \cite{DiasUnpub}.}  We recommend Ref.~\cite{Sasaki:1995aw} for a thorough derivation of these quantities.

We start with the action for $N_f$ scalar fields $\phi^I$, given by
\begin{equation}
S = \int \ud^4 x \sqrt{-g} \left[ -\frac{1}{2} G_{IJ} \pd_\mu \phi^I \pd^\mu \phi^J - V(\f^I) \right].
\label{eqn:gen_action}
\end{equation}
Again, we assume that Greek indices $\alpha \in [0,\dots,3]$ are for spacetime, upper-case Latin letters index the number of fields, $I \in [1,\dots,N_f]$, and lower-case Latin letters index three-space, $i \in [1, \dots, 3]$.  The action~\eqref{eqn:action} is a special case of Eq.~\eqref{eqn:gen_action}, with $G_{IJ} = \delta_{IJ}$.  To change the equations in \S\ref{sect:pert_review} to reflect the general field-space metric, we follow the typical procedure of replacing partial derivatives with respect to the fields with covariant derivatives.

Varying Eq.~\eqref{eqn:gen_action} with respect to the fields gives the background equation of motion
\begin{eqnarray}
\frac{D\f^I}{\ud t} + 3H \dot\f^I + G^{IJ} V_{; J} = 0 ,
\label{eqn:back_gen}
\end{eqnarray}
where we have assumed that the background fields are homogeneous, $\partial_i \phi_J = 0$.
In Eq.~\eqref{eqn:back_gen} we have used the covariant differential~\cite{Sasaki:1995aw}
\begin{equation}
D \f^I = \ud \f^I + \Gamma^I_{JK}\dot{\f^J}\ud\f^K ,
\end{equation}
with the field-space Christoffel symbols
\begin{equation}
\Gamma^I_{JK} = \frac{1}{2} G^{IL}\left( G_{LJ, K} + G_{LK,J} - G_{JK, L} \right) ,
  \label{eqn:XXX}
\end{equation}
which gives the covariant derivative $V_{;I} \equiv D V / d \f^I = \pd V/\pd \phi^I$, because $V$ is a scalar.

Following the treatment in Section~\ref{sect:pert_review}, it is convenient to express the equations of motion for the field perturbations $\delta \phi^I_\bk$ in spatially-flat gauge in terms of the generalized trajectory velocity $\dot \phi_0$, which can be extended from Eq.~\eqref{eqn:clock} to
\begin{equation}
{\dot\f}_0^2 \equiv G_{IJ} {\dot\f}^I {\dot\f}^J .
  \label{eqn:XXX}
\end{equation}
From $\dot \phi_0$, we define the adiabatic unit vector in the kinematic basis~\cite{Gordon:2000hv,GrootNibbelink:2001qt} as $\omega^I \equiv \dot \phi^I / \dot \phi_0$, which projects vectors along the adiabatic direction in the generalized field-space.  Finally, the mode equations in spatially-flat gauge read~\cite{Sasaki:1995aw}
\begin{equation}
\frac{D^2 }{\ud t^2} \delta \phi_\bk^I + 3H \, \frac{D }{\ud t} \delta \phi_\bk^I + \frac{k^2}{a^2} \delta \phi^I_\bk + {C^I}_J \delta \phi^J_\bk = 0,
\end{equation}
which is a simple generalization of Eq.~\eqref{eqn:dphi_mode}, with the mass matrix
\begin{equation}
{C^I}_J = G^{IK}V_{;K;J}
- {\dot\f_0}^2 \, {R^I}_{JKL} \omega^K \omega^L
+ 2\epsilon \frac{H}{\dot\f_0} \, \left(\omega^I V_{;J} + \omega_J V^{;I} \right)
+ 2\epsilon(3-\epsilon) H^2 \, \omega^I \omega_J ,
\end{equation}
where ${R^I}_{JKL}$ is the field-space Riemann tensor, built from $G_{IJ}$.  Again, we note that index contraction implies summation with respect to the field-space metric, $X_I Y^I = G_{IJ} X^I Y^J$.

Quantizing the modes using the mode-matrix $\delta \phi_I = {\psi_{I}}^J a_J$ proceeds as in Section~\ref{ssect:mode} and the adiabatic power spectrum $\PR$ is identical to Eq.~\eqref{eqn:pad}, except contracting with respect to $G_{IJ}$.  The isocurvature directions should again be found by Gram-Schmidt orthogonalization, except now implemented in the curved field-space, and the spectra $\Ppnad$ and $\Pent$ can be found as in Eqs.~\eqref{eqn:ppnad}~and~\eqref{eqn:pent}.  The projected isocurvature spectrum $\PS$ in Eq.~\eqref{eqn:piso} can instead be built by replacing the summation over the isocurvature directions by contraction with respect to the isocurvature directions of the field-space metric, $\hat G_{IJ}$, after transforming to the kinematic basis.

\section{\MMC usage}
\label{app:usage}

\MMC is publicly available at \url{www.modecode.org} and is released with a Modified BSD License.  \MMC has been implemented in a mix of Fortran 95/2000 and has been thoroughly checked on Mac OS and Linux systems with the freely-available GNU Fortran compiler (version 4.6.3+) and with Intel Fortran (version 14.0.2), with the Intel compiler yielding significant improvements in speed.  There are no external dependencies necessary to use \MMC.

We include a driver file \code{multimodecode\_driver.f90} that contains the basic structure needed to explore the predictions of a model.  The driver has many important routines for calculating the power spectrum and outputting the results.  The file \code{parameters\_multimodecode.txt} contains runtime parameters that are often changed between subsequent runs.  The parameters are listed in Fortran namelists, so a change here does not require the user to recompile the whole code.  There are a mix of basic and advanced parameters available and we describe them here.

The \code{\&init} namelist holds important parameters related to initializing the program, the choice of inflationary potential $V$, and program output:
\begin{lstlisting}
&init
  num_inflaton = 10
  potential_choice = 1
  vparam_rows = 4
  slowroll_infl_end = .true.
  instreheat = .false.
/
\end{lstlisting}
The number of fields is set with the variable \code{num\_inflaton} and an indentifying number is chosen for the potential with \code{potential\_choice}.  The currently available potentials are listed in the routine \code{pot(phi)} in \code{modpk\_potential.f90}, including the multifield $N_f$-quadratic \cite{Linde:1983gd,Liddle:1998jc,Dimopoulos:2005ac}, $N_f$-monomial inflation with $V \sim \lambda_I |\phi_I|^n$ \cite{Price:2014ufa}, two-field hybrid inflation \cite{Linde:1993cn,Copeland:1994vg}, a product of exponentials \cite{Liddle:1998jc}, and the multifield generalization of the hyperbolic-tangent step potential in Ref.~\cite{Adams:2001vc}, which was used in Section~\ref{ssect:case}.  Adding a new potential is as easy as providing the potential $V$ and its derivatives in \code{modpk\_potential.f90} with a new value for \code{potential\_choice}.  The array \code{vparams} contains information passed to the potential function (\emph{e.g.}, masses and couplings) and has dimensions (\code{vparam\_rows})$\times$(\code{num\_inflaton}).  The values for \code{vparams} are set in \code{parameters\_multimode.txt} in the namelist \code{\&params}; variables related to the pivot scale \code{N\_pivot} and \code{k\_pivot} are also set here.

The user can change the conditions for inflation to end by varying \code{slowroll\_infl\_end}, which when set to \code{.true.} evolves the background fields until $\epsilon=1$.  If you do not want this to be the ending criterion, then set \code{slowroll\_infl\_end=.false.} and adapt the subroutine \code{alternate\_infl\_end} in \code{modpk\_odeint.f90} to change the ending condition.  Furthermore, with \code{instreheat=.true.} $N_*$ becomes a derived parameter by requiring inflation to thermalize immediately after the end of inflation with $w=1/3$, as in Ref.~\cite{Easther:2011yq}.

We use the \code{\&analytical} namelist to set the options for calculating the power spectrum, either using the $\delta N$ calculations of \S\ref{sect:dN} or evaluating the full mode equations of \S\ref{ssect:mode} or both.
\begin{lstlisting}
&analytical
  use_deltaN_SR = .true.
  use_horiz_cross_approx = .false.
  evaluate_modes = .true.
  get_runningofrunning = .false.
/
\end{lstlisting}
With \code{use\_deltaN\_SR=.true.} \MMC will calculate the $\delta N$ observables at the pivot scale (as given in namelist \code{\&params}) and if \code{use\_horiz\_cross\_approx=.true.}, it will ignore the contribution to $N_{,I}$ and $N_{,IJ}$ from the end-of-inflation surface via the horizon crossing approximation (HCA)~\citep{Vernizzi:2006ve,Kim:2006te}.  Setting \code{evaluate\_modes=.false.} will make the program only solve the background equations of motion, relying on the $\delta N$ calculations to obtain the spectra.  If \code{get\_runningofrunning=.true.}, then the derivative of $\alpha_s$ with respect to $\ln k$ will be calculated by a five-point stencil finite difference method, which requires two additional calls to the code that solves the mode equations, significantly slowing down the speed of the program.

The way by which the initial conditions are chosen for a given simulation depends on the namelist \code{\&ic\_sampling\_nml} namelist.
\begin{lstlisting}
&ic_sampling_nml
  ic_sampling = 1
  numb_samples = 1
  energy_scale = .1
  save_iso_N = .false.
  N_iso_ref = 55
/
\end{lstlisting}
The variable \code{ic\_sampling} controls the main behavior of the initial conditions' prior probability and is set to an identifying number defined in the file \code{modpk\_icsampling.f90} with the \code{ic\_samp\_flags} type.  The currently available values are below:
\begin{lstlisting}
  type :: ic_samp_flags
    integer :: reg_samp = 1
    integer :: eqen_samp = 2
    integer :: slowroll_samp = 3
    integer :: iso_N = 6
  end type
\end{lstlisting}
Invoke each case by setting \code{ic\_sampling} equal to the desired number.  The functionality of each of these cases is:
\begin{itemize}

  \item \code{reg\_samp} (regular sampling): the simple case of setting a multifield initial condition and calculating the power spectrum.  The initial conditions for the fields are set as the variable \code{phi\_init0} in the \code{\&params} namelist and the fields' velocities are assumed to be initially in slow-roll.

  \item \code{eqen\_samp} (equal-energy sampling): quasi--equal-area sampling of a phase-space surface with same initial energy, as in Refs~\cite{Easther:2013bga,Easther:2013rva,Easther:2014zga}.  Set the initial energy with the \code{energy\_scale} variable in units of $\mpl$, where $\mpl^2 = (8 \pi G)^{-1} =1$.  To also record and output the background field values as the background reaches $N_e=$ \code{N\_iso\_ref} set \code{save\_iso\_N=.true.}.  The prior ranges for the fields and velocities are set in the namelist \code{\&priors}.

  \item \code{slowroll\_samp} (slow-roll sampling): choose initial conditions in field space and set velocities by the slow-roll condition.  The prior ranges for the fields are again chosen in \code{\&priors}.

  \item \code{iso\_N} (sampling $e$-fold surface): uniformly samples the surface $N_e = \sum_I \phi_I^2/2p$ for $N_f$-monomial inflation, as in Refs~\cite{Frazer:2013zoa,Easther:2013rva}.  Set the value for $N_e$ with \code{N\_iso\_ref} in the \code{\&ic\_sampling} namelist.

\end{itemize}
To implement a different initial conditions measure or sampling behavior, add a new identifier for  the \code{ic\_samp\_flags} type.  The initial conditions are set in the routine \code{get\_ic} in \code{modpk\_icsampling.f90} and you will need to implement your sampling technique here, following the examples in the code.

Similarly, the prior probabilities on the \code{vparams} array that defines the potential are set through the \code{\&param\_sampling\_nml} namelist.
\begin{lstlisting}
&param_sampling_nml
  param_sampling = 1
  use_first_priorval = .true.
  vp_prior_min(1,:) = -14
  vp_prior_max(1,:) = -12
  varying_N_pivot = .false.
/
\end{lstlisting}
As with the initial conditions sampler, different behaviors are chosen by setting the variable \code{param\_sampling} to a unique integer as specified in the \code{param\_samp\_flags} type in \code{modpk\_icsampling.f90}.
\begin{lstlisting}
type :: param_samp_flags
  integer :: reg_constant = 1
  integer :: unif_prior = 2
  integer :: log_prior = 3
end type
\end{lstlisting}
Again, invoke each case by setting \code{param\_sampling} to the desired number.  To vary the number of $e$-folds after the pivot scale leaves the horizon, set \code{varying\_N\_pivot=.true.} and set the limits on the prior on $N_*$ in the \code{\&priors} namelist, where we have assumed a uniform prior.  Note that this is overridden if \code{instreheat=.true.}.

To use a different prior on the \code{vparams} array, add a new integer to the \code{param\_samp\_flags} type and change the routine \code{get\_vparams} in \code{modpk\_icsampling.f90}.
The default behavior is:
\begin{enumerate}

  \item \code{reg\_constant} (regular, constant parameters): the \code{vparams} array is kept constant, as specified in the \code{\&params} namelist.

  \item \code{unif\_prior} (uniform prior probability): each column in \code{vparams} is chosen with a uniform prior between \code{vp\_prior\_min} and \code{vp\_prior\_max}.  If \code{use\_first\_priorval=.true.}, then the first entry in the priors are used for all the columns.

  \item \code{log\_prior} (logarithmic prior probability): the columns of a dummy array $\alpha_{IJ}$ are chosen with a uniform prior according to the priors in the namelist.  The columns in the \code{vparams} are then set by \code{vparams}=$10^{\alpha_{IJ}}$.

\end{enumerate}

The \code{\&params} namelist is used to set the \code{vparams} array, the fields' initial conditions \code{phi\_init0}, the pivot scale \code{k\_pivot}, and the number of $e$-folds between horizon exit and the end of inflation for the pivot scale by \code{N\_pivot}.
\begin{lstlisting}
&params
  N_pivot = 55.0
  k_pivot = 0.002
  dlnk = 0.4
  phi_init0 = 10.0 10.0 10.0 10.0 10.0 10.0 10.0 10.0 10.0 10.0
  vparams(1,:) = -11.4 -11.0 -10.8 -10.6 -10.5 -10.4 -10.3 -10.2 -10.1 -9.9
  vparams(2,:) = 8.8  7.3  6.0  4.8  3.6  2.6  1.7  1.0  5.6  0.16
  vparams(3,:) = 1e-3 1e-3 1e-3 1e-3 1e-3 1e-3 1e-3 1e-3 1e-3 1e-3
  vparams(4,:) = 1e-2 1e-2 1e-2 1e-2 1e-2 1e-2 1e-2 1e-2 1e-2 1e-2
/
\end{lstlisting}
These are all default values and may be overridden with different choices of sampling, as mentioned above.  The variable \code{dlnk} is the difference in $k$-space that is used when we evaluate the pivot-scale observables from the mode equations via finite difference.  All scales are in units of $\impc$ and fields are in $\mpl$.

The prior probability ranges for the sampling of the fields' initial values and $N_*$ are in the \code{\&priors} namelist, which is relatively self-explanatory.
\begin{lstlisting}
&priors
  phi0_priors_min = 2.0 2.0
  phi0_priors_max = 30.0 30.0
  dphi0_priors_min = -1.262e0 -1.262e0
  dphi0_priors_max =  1.262e0 1.262e0
  N_pivot_prior_min = 30
  N_pivot_prior_max = 70
/
\end{lstlisting}

\MMC defaults to obtaining only the pivot-scale observables by taking the numerical derivative of the power spectra at $k_*$.  However, the full power spectra can also be solved for and provide the full description of the model's predictions over the scales of interest, as in \S\ref{ssect:case}.  We control this through the \code{\&full\_pk} namelist.
\begin{lstlisting}
&full_pk
  calc_full_pk = .false.
  steps = 300
  kmin = 1.0e-4
  kmax = 1.0e0
/
\end{lstlisting}
The variable \code{calc\_full\_pk} options the calculation of the full $\mathcal P(k)$.  The program will interpolate between a number of points, given by the variable \code{steps}, between the scales \code{kmin} and \code{kmax}.

\MMC is able to save and output a significant amount of data for later analysis.  However, since this has an obvious affect on the speed of the code, the amount and verbosity of this output can be specified by the attributes of the \code{out\_opt} instance of the \code{print\_options} type in the \code{print\_out} namelist.
\begin{lstlisting}
&print_out
  out_opt%modpkoutput = .true.
  out_opt%output_reduced = .true.
  out_opt%output_badic =.false.
  out_opt%save_traj = .true.
  out_opt%fields_horiz = .false.
  out_opt%fields_end_infl = .false.
  out_opt%spectra = .false.
  out_opt%modes = .false.
/
\end{lstlisting}
If you want nothing to write to stdout (the terminal, usually), then set \code{modpkoutput=.false.}; if less output is requested, then set \code{output\_reduce=.true.}.  If \code{output\_badic=.false.}, then any set of parameters that do not lead to a successfully inflating universe are discarded and ignored; otherwise, they will be saved and output into the data files with dummy values for their spectra.  The remaining attributes of \code{out\_opt} controls what cosmological quantities are printed to file in addition to the pivot scale observables: \code{save\_traj} records the background trajectory as a function of $N_e$; \code{fields\_horiz} saves the field values as the pivot scale crossing the horizon; \code{fields\_end\_infl} saves the field values at the end of inflation; \code{spectra} will record the superhorizon power spectra as a function of $N_e$; and \code{modes} prints all the mode functions during the entire evolution.  Setting \code{modes=.true.} consequently results in a lot of output.

The output of \MMC is saved in comma-delimited CSV files, with the first row corresponding to a header that names each column.  Subsequent rows correspond to different samples of the same model with different parameters.  To change the output, simply find the point where the header is written in the code, add a new column(s), and print out the desired quantity in the correct order.

Finally, some technical options are controllable via the \code{\&technical} namelist, by changing the attributes of the \code{tech\_opt} instance of the \code{tech\_options} type.  In particular, the choice of numerical integration scheme can be changed by the \code{tech\_opt\%use\_dvode\_integrator} flag.  Setting this to \code{.true.} will invoke a backwards-difference formula method, which is suitable for stiff problems, while \code{.false.} will use a fourth-order Runge-Kutta integrator.  Various accuracy settings for the integrators can also be controlled in this namelist, with the global behavior set by the variable \code{accuracy\_setting}=$-1,1,2$.  Using 1 sets a minimal amount of accuracy, which we find is suitable for calculating the adiabatic power spectrum for simple models.  Increasing this to 2, increases the accuracy and, in particular, increases the accuracy as the evolution moves out of slow-roll, which we find is necessary to obtain resolved isocurvature spectra, as in Fig.~\ref{fig:pk_compare}.  Lastly, you can override our choices for the absolute and relative error tolerances by setting \code{accuracy\_setting=-1} and manually changing the remaining attributes of \code{tech\_opt} in the namelist, which is self-explanatory.

\bibliographystyle{JHEP}

\bibliography{references}

\end{document}